\documentclass[pre,reprint]{revtex4-1}
\usepackage{bm,epsfig,graphicx,color,amssymb,amsmath}

\begin{document}

\title{Renormalization of Optical Transition Strengths in Semiconductor Nanoparticles due to Band Mixing}

\author{Kirill A. Velizhanin}
\email{kirill@lanl.gov}
\affiliation{Theoretical Division, Los Alamos National Laboratory, Los Alamos, New Mexico 87545, USA}

\begin{abstract}
Unique optical properties of semiconductor nanoparticles (SN) make them very promising in the multitude of applications including lasing, light emission and photovoltaics. In many of these applications it is imperative to understand the physics of interaction of electrons in a SN with external electromagnetic fields on the quantitative level. In particular, the strength of electron-photon coupling determines such important SN parameters as the radiative lifetime and absorption cross section. This strength is often assumed to be fully encoded by the so called Kane momentum matrix element. This parameter, however, pertains to a bulk semiconductor material and, as such, is not sensitive to the quantum confinement effects in SNs. In this work we demonstrate that the quantum confinement, via the so called band mixing, can result in a significant suppression of the strength of electron interaction with electromagnetic field. Within the envelope function formalism we show how this suppression can be described by introducing an effective energy-dependent Kane momentum. Then, the effect of band mixing on the efficiencies of various photoinduced processes can be fully captured by the conventional formulae (e.g., spontaneous emission rate), once the conventional Kane momentum is substituted with the renormalized energy-dependent Kane momentum introduced in here. As an example, we evaluate the energy-dependent Kane momentum for spherical $\rm{PbSe}$ and $\rm{PbS}$ SNs (i.e., quantum dots) and show that neglecting band mixing in these systems can result in the overestimation of absorption cross sections and emission rates by a factor of $\sim$2. 
\end{abstract}

\maketitle

\section{Introduction}

Unique properties of photoinduced electronic processes in semiconductor
nanoparticles (SN), as compared to bulk semiconductors, is what makes
SNs (e.g., quantum dots, nanorods) so promising in the multitude of
applications including photovoltaics, photonics, lasing and light-emitting
devices \cite{Ip-2012-577,Chuang-2014-796,TenCate-2015-174,Schaller-2004-186601,Klimov-2007-441,Kim-2008-4513,Bae-2013-3661}. As follows from the term ``photoinduced'', such processes
are all initiated by an optical absorption of a photon (or multiple
photons) resulting in electronic transitions within a SN. The efficiency
of initiation of such processes is thus directly determined by the
strength of interaction between external electromagnetic fields (EMF)
and SN electronic transitions. The same interaction does often further
affect the electronic dynamics of initially photoexcited charge carriers.
For example, the electron-EMF coupling determines the rate of F\"orster
resonance energy transfer between nearby SNs. Photon absorption, as
well as spontaneous and stimulated photon emissions, are main processes
determining the efficiency of lasing. Finally, the lifetime of a charge
carrier is affected by the electron-EMF coupling via the radiative
recombination rate. This is especially true for high-quality SNs where the quantum yield of photoluminescence can approach
100\%, implying that the observable lifetime of charge carriers approaches
the inverse radiative recombination rate \cite{Qu-2002-2049,Chen-2013-445}.

The strength of coupling between electronic transitions in SNs
and EMF is often expressed via the so called Kane momentum matrix
element or simply Kane momentum, $P=\hbar\langle u_{c}|\nabla|u_{v}\rangle$,
where $u_{c}$ and $u_{v}$ are bulk Bloch functions corresponding
to conduction and valence band edges, respectively \cite{YuCardona-Fundamentals-1999}.
For example, approximating an SN as a simple two-level quantum emitter
in vacuum results in the following expression for the rate of radiative
recombination (in Gaussian units) \cite{Efros-1993-10005,Liu-2010-14860}
\begin{equation}
k_{r}=\frac{4e^{2}EP^{2}}{3\hbar^{2}m^{2}c^{3}},\label{eq:kr_P}
\end{equation}
where $e=|e|$ and $m$ are the magnitude of the charge and the mass of free electron,
respectively. Speed of light in vacuum and Planck constant are denoted by $c$ and $\hbar$, respectively. Energy
of the lowest electronic transition, $E$, could be significantly
larger than the bandgap energy of the corresponding bulk semiconductor,
$E_{g}$, due to quantum confinement effects in SNs. Cross sections of absorption
and stimulated emission are linearly related to $k_{r}$ via Einstein
coefficients \cite{Hillborn-1982-982}, and so are also proportional
to $P^{2}$. For simplicity, we disregard the effect of dielectric screening within the SN and the surrounding material (e.g., solvent) on the electron-EMF coupling.  These effects can always be straightforwardly accounted for, as described in Ref.~\cite{Wehrenberg2002-10634}.

Representing a wavefunction of a conduction (valence) band electron in SN as a product of a slowly varying \emph{envelope} function and the Bloch function $u_c$ ($u_v$) \cite{YuCardona-Fundamentals-1999,Haug-Koch-2009}, one can relate the Kane momentum to the transition dipole moment of the lowest transition as $d=\left|\langle u_{c}|e{\bf r}|u_{v}\rangle\right|=\frac{e\hbar P}{mE_{g}}$
\cite{Rosencher-Vinter-2002,Haug-Koch-2009}, as discussed in more detail in Sec.~\ref{sec:LengthGauge}. Using this transition
dipole moment, the radiative recombination rate can now be written as \cite{Landafshitz-4}
\begin{equation}
k_{r}=\frac{4E^{3}d^{2}}{3\hbar^{4}c^{3}}.\label{eq:kr_d}
\end{equation}
Since $P$ and $d$ are bulk properties, and as such are independent of specific value of transition energy $E$, Eqs.~(\ref{eq:kr_P}) and (\ref{eq:kr_d})
imply $k_{r}\propto E$ in and $k_{r}\propto E^{3}$, respectively. Therefore, Eqs.~(\ref{eq:kr_P}) and (\ref{eq:kr_d}) do contradict each other.
This apparent contradiction in scaling of $k_{r}$ with respect to
the energy of the lowest transition can often be ignored for wide-bandgap SNs (e.g., $\rm{CdSe}$ or $\rm{CdS}$), where $E\approx E_{g}$ for
typical experimentally relevant SN sizes. However, $E/E_{g}$ could
be as high as $\sim3-5$ in small narrow-gap SNs
due to strong quantum confinement. Under these conditions, the issue
of contradictory energy scaling in Eqs.~(\ref{eq:kr_P}) and (\ref{eq:kr_d})
becomes critical and has to be resolved.

\begin{figure}
\includegraphics[width=8.7cm]{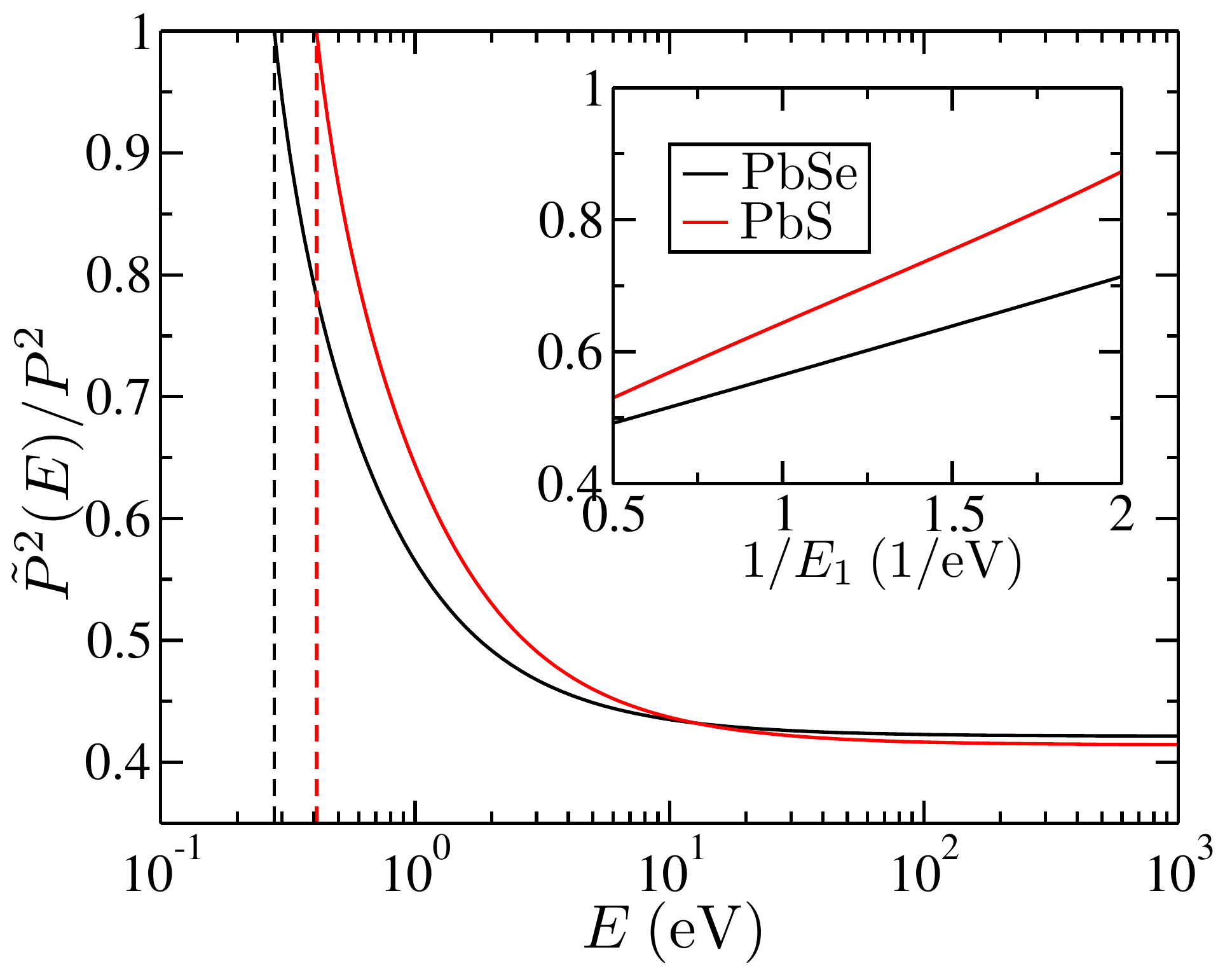}\caption{\label{fig:P_E}
Normalized square of the effective energy-dependent Kane momentum for the lowest-energy transition in $\rm{PbSe}$ (black) and $\rm{PbS}$ (red) QDs as a function of transition energy, $E$. Vertical dashed lines represent bulk bandgap energies. Inset shows the same data within the experimentally relevant energy range.
}
\end{figure}

The main goal of the present paper is to carefully investigate the effect of quantum confinement in narrow-gap SNs on the strength of coupling between electronic transitions and EMF. Working within the envelope function approximation ($k\cdot p$ model) \cite{YuCardona-Fundamentals-1999,Haug-Koch-2009}, we demonstrate that the requirement of gauge invariance of the $k\cdot p$ model disallows the energy independence of $P$ and $d$ in Eqs.~(\ref{eq:kr_P}) and (\ref{eq:kr_d}). Specifically, it is demonstrated how the band mixing (or band coupling) \cite{Efros-1998-7120,Voon-Willatzen-2009} in SNs \emph{renormalizes} bulk values of $P$ and $d$ by making them effectively energy-dependent. Such renormalized
energy-dependent parameters, $\tilde{P}(E)$ and $\tilde{d}(E)$,
do yield exactly the same $k_{r}(E)$, once substituted into Eqs.~(\ref{eq:kr_P})
and (\ref{eq:kr_d}) instead of bulk $P$ and $d$, respectively.
This further implies that $E^{2}\tilde{d}^{2}(E)$ has to be linearly
proportional to $\tilde{P}^{2}(E)$ as a function of $E$, and we
show how this relation follows exactly from the gauge invariance of
the $k\cdot p$ model. Using a simple one-dimensional two-band Kane
model, we show how different regimes of quantum confinement affect the strength of the electron-EMF interaction. We further
discuss a specific experimentally relevant example of $\rm{PbSe}$ and $\rm{PbS}$ spherical
SNs, i.e., quantum dots (QD), and show that even though renormalized
$\tilde{P}(E)$ and $\tilde{d}(E)$ are both energy-dependent, $\tilde{P}(E)$
dependence on energy is much weaker than that of $\tilde{d}(E)$,
so the former could be represented by a simple linear fit (or even
constant) within the experimentally relevant range of QD sizes. This
result is consistent with previous experimental and theoretical observations
\cite{Allan-2004-245321,Moreels-2009-3023}. Fig.~\ref{fig:P_E}
plots the renormalized value of the Kane momentum for the lowest-energy optical transition
in $\rm{PbSe}$ and $\rm{PbS}$ QDs, as evaluated within a suitable $k\cdot p$ model
\cite{Kang-1997-1632}, as a function of the transition energy.
%
%
As is seen, the square of the renormalized Kane momentum does change
by $\sim$$40\%$ in the experimentally relevant energy window
($E=0.5-2$ eV). This result can be directly used to evaluate radiative
lifetimes in $\rm{PbSe}$ and $\rm{PbS}$ QDs.

The paper is organized as follows. The microscopic Hamiltonian for
an electron in a bulk semiconductor in the presence of EMF is discussed
in Sec.~\ref{sec:mic_Ham}. Sec.~\ref{sec:envelope} focuses on the envelope function Hamiltonian suitable for description of electronic structure and optical transitions in SNs. The effect of band mixing
on the strength of optical transitions in a one-dimensional two-band
Kane model is considered in Sec.~\ref{sec:2b_Kane}. In particular,
band mixing in the continuos and spatially confined Kane models is
described in Secs.~\ref{sub:Cont_Kane} and \ref{sub:conf_Kane},
respectively. Practically relevant example of lead chalcogenide QDs
is discussed in Sec.~\ref{sec:lead_chalc}. Sec.~\ref{sec:Discussion}
reveals the necessity of accounting for band mixing when evaluating
the intensities of optical transitions, and shows how disregarding band
mixing often results in breaking the gauge invariance of the $k\cdot p$
model. Sec.~\ref{sec:Conclusion} concludes.

\section{Microscopic Hamiltonian\label{sec:mic_Ham}}

A single-particle Hamiltonian for an electron in a bulk semiconductor
in the absence of external EMF reads as (vectors are written in bold)
\begin{equation}
H_{0}({\bf r},{\bf p})=\frac{{\bf p}^{2}}{2m}+V({\bf r},{\bf p}),\label{eq:micHam}
\end{equation}
where ${\bf r}$ is the position vector, and ${\bf p}=-i\hbar\nabla$ is the momentum operator. The operator
of potential energy, $V({\bf r},{\bf p})$, is periodic over the semiconductor
lattice and can, in general, have contributions linear in momentum
${\bf p}$ due to spin-orbit interaction.

The direct evaluation of the radiative recombination rate from the microscopic Hamiltonian is somewhat tedious
since it requires quantization of EMF. Optical absorption is simper to assess as the EMF can be treated classically.
Therefore, in this paper we focus on the strength of coupling of electronic
transitions to the classical EMF with the understanding that renormalized
$\tilde{P}(E)$, obtained from such analysis, is exactly the same
$\tilde{P}(E)$ that enters Eq.~(\ref{eq:kr_P}), by virtue of Einstein coefficients \cite{Hillborn-1982-982}.

Interaction with a time-dependent classical EMF is introduced into Eq.~(\ref{eq:micHam}) via the so called minimal-coupling procedure: (i)
the momentum operator is ``elongated'' by the vector potential, ${\bf p}\rightarrow{\bf p}+\frac{e}{c}{\bf A}({\bf x},t)$,
and (ii) the scalar potential $\varphi({\bf r},t)$ is added, so that the resulting Hamiltonian is (in Gaussian
units) \cite{Landafshitz-2}
\begin{equation}
H({\bf r},{\bf p},t)=H_{0}\left({\bf r},{\bf p}+\frac{e}{c}{\bf A}({\bf r},t)\right)-e\varphi({\bf r},t).\label{eq:Hem}
\end{equation}
The gauge transformation of the vector and scalar potentials is given by
\begin{gather}
{\bf A}({\bf r},t)\rightarrow\tilde{{\bf A}}({\bf r},t)={\bf A}({\bf r},t)+\nabla f({\bf r},t),\nonumber \\
\varphi({\bf r},t)\rightarrow\tilde{\varphi}({\bf r},t)=\varphi({\bf r},t)-\frac{1}{c}\frac{\partial f({\bf r},t)}{\partial t},\label{eq:gauge_trans}
\end{gather}
where $f({\bf r},t)$ is an arbitrary real scalar field. It is
straightforward to demonstrate that if an arbitrary wavefunction $\Psi({\bf r},t)$
is the solution of the time-dependent Schr\"odinger equation with Hamiltonian
(\ref{eq:Hem}), then $\tilde{\Psi}({\bf r},t)=e^{-i\frac{e}{\hbar c}f({\bf r},t)}\Psi({\bf r},t)$
is the solution of the Schr\"odinger equation with the same Hamiltonian
where the vector and scalar potentials are transformed according to
Eq.~(\ref{eq:gauge_trans}). Therefore, the gauge transformation
of the vector and scalar potentials is exactly equivalent to a certain
unitary transformation of the wavefunction, which obviously does not
affect any measurable quantities, e.g., charge density $\left|\Psi({\bf x},t)\right|^{2}$
or rates of photoinduced transitions between the states. The dynamics of the system,
encoded by Hamiltonian (\ref{eq:Hem}), is thus \emph{gauge invariant}.
In what follows, we will loosely refer to Hamiltonians of the type
given by Eq.~(\ref{eq:Hem}) as gauge invariant, even though, strictly
speaking, the Hamiltonian operator does not transform into itself
upon substitution (\ref{eq:gauge_trans}). 

Although physical observables are gauge invariant, certain calculations
could be simplified by an appropriate choice of gauge. Two specific
gauge choices are of particular convenience when describing the interaction
of SNs with EMF. Starting from an arbitrary gauge it is always possible
to choose $f({\bf r},t)=-c\int_{t_{0}}^{t}dt'\,\varphi({\bf r},t')$
in Eq.~(\ref{eq:gauge_trans}) so that the scalar potential vanishes exactly upon the gauge transformation, resulting in
\begin{equation}
{\bf A}({\bf r},t)={\bf A}(t)={\bf A}_{0}\sin(\omega_{0}t);\,\varphi({\bf r},t)\equiv0,\label{eq:gauge1}
\end{equation}
where it is assumed that the electronic transitions in SN are driven
by the oscillatory EMF of angular frequency $\omega_{0}$. The constant vector
amplitude of the EMF vector potential is ${\bf A}_{0}$. The dependence
of the vector potential on ${\bf r}$ is neglected as it is typically
very weak for, e.g., electromagnetic waves of relevant frequencies.
For the free electromagnetic wave, the gauge (\ref{eq:gauge1}) coincides
with the Coulomb (transverse) gauge. If EMF is weak, we can expand the microscopic Hamiltonian (\ref{eq:Hem}) with respect to the vector potential and retain only up to linear terms. If we further neglect the spin-orbit interaction, the resulting electron-EMF coupling operator is proportional to ${\bf A}_0\cdot{\bf p}$. Since momentum ${\bf p}$ is related to velocity, gauge~(\ref{eq:gauge1}) is often called the \emph{velocity} gauge.

The second convenient choice of gauge can be obtained from Eq.~(\ref{eq:gauge1})
by performing the gauge transformation, Eq.~(\ref{eq:gauge_trans}),
with $f({\bf r},t)=-({\bf r}\cdot{\bf A}_{0})\sin(\omega_{0}t)$. This produces 
\begin{equation}
{\bf A}({\bf r},t)\equiv0;\,\varphi({\bf r},t)=\frac{\omega_{0}}{c}({\bf r}\cdot{\bf A}_{0})\cos(\omega_{0}t).\label{eq:gauge2}
\end{equation}
Substituting this expression into Eq.~(\ref{eq:Hem}) one can see that the electron-EMF coupling operator is linearly proportional to ${\bf r}$. Hence, gauge (\ref{eq:gauge2})  is often referred to as the \emph{length} gauge. 

The convenience of the two gauge choices, Eqs.~(\ref{eq:gauge1}) and (\ref{eq:gauge2}), stems from the fact that electronic transitions in SNs are coupled to EMF solely through either vector or scalar potential for the velocity and length gauge, respectively.
It has to be emphasized here, however, that we are able to make the
vector potential vanish in Eq.~(\ref{eq:gauge2}) only because we
neglected its coordinate dependence in Eq.~(\ref{eq:gauge1}). More
generally, one can make a vector potential vanish only locally in space by choosing a gauge.

The gauge invariance is absolutely necessary for the accurate and
consistent treatment of interaction of charge carriers with EMF.
For example, neglecting the vector potential in Eq.~(\ref{eq:Hem}),
and thus breaking gauge invariance, can result in unphysical gauge-dependent
rates of photoinduced transitions, which could be made as large as
desired by a choice of gauge. An example of such behavior is provided
in Sec~\ref{sub:velocity_gauge}.

\section{Envelope function approximation\label{sec:envelope}}

To analyze the electronic structure of SNs, it is convenient to not directly
use Hamiltonian (\ref{eq:micHam}), but to transform
it first to the envelope function representation \cite{YuCardona-Fundamentals-1999,Haug-Koch-2009}.
To this end, we find the eigenfunctions $\chi_{n{\bf k}}({\bf r})$
of the bulk Hamiltonian (\ref{eq:micHam}) at a specific (typically
high symmetry) point ${\bf k}$ of the Brillouin zone
\begin{equation}
\chi_{n{\bf k}}({\bf r})=\frac{1}{\sqrt{V}}e^{i{\bf k}\cdot{\bf r}}u_{n{\bf k}}({\bf r}).\label{eq:bwf_basis}
\end{equation}
Here, index $n$ enumerates electronic bands, and Bloch functions
$u_{n{\bf k}}({\bf r})$ are periodic over the lattice. These eigenfunctions
can now be used to expand an arbitrary time-dependent wavefunction $\Psi({\bf r},t)$ as
\begin{equation}
\Psi({\bf r},t)=\sum_{n}\phi_{n}({\bf r},t)\chi_{n{\bf k}}({\bf r}),\label{eq:Psi_micr}
\end{equation}
where $\phi_{n}({\bf r},t)$ are the so called envelope functions
that are assumed to vary slowly over the size of a single unit cell.
Since the basis functions $\chi_{n{\bf k}}({\bf r})$ are time-independent,
it is possible to write down an effective low-energy Hamiltonian that
acts only within a space of envelope functions. This is done by multiplying
the Hamiltonian (\ref{eq:micHam}) by the basis set functions from
both sides and then averaging over the unit cell (uc)
\begin{equation}
H_{0,nm}({\bf r},{\bf p})=\langle\chi_{n{\bf k}}|H_{0}({\bf r},{\bf p})|\chi_{m{\bf k}}\rangle_{{\rm uc}}.\label{eq:Henv}
\end{equation}
Hamiltonian $H_{0,nm}$ is a matrix of operators that is at most quadratic
with respect to ${\bf p}$. Typically, such Hamiltonians include terms
proportional to the dot product of ${\bf k}$ and ${\bf p}$, hence
the term $k\cdot p$ Hamiltonian. These operators act only within
the space of envelope functions, so that a time-dependent Schr\"odinger
equation for the wavefunction in Eq.~(\ref{eq:Psi_micr}) can now
be expressed as
\begin{equation}
\sum_{m}H_{0,nm}({\bf r},{\bf p})\phi_{m}({\bf r},t)=i\hbar\frac{\partial\phi_{n}({\bf r},t)}{\partial t}.\label{eq:kp_inf}
\end{equation}

The obtained Hamiltonian $H_{0,nm}({\bf r},{\bf p})$ is still not
quite suitable for practical calculations since the summation over
$m$ in Eq.~(\ref{eq:kp_inf}) runs over an infinite number of electronic
bands. The number of bands can be reduced to a finite number of \emph{relevant}
ones by means of an approximate canonical transformation that decouples
the so called far-bands from the experimentally relevant ones (e.g.,
highest valence and lowest conduction bands) within the second-order
perturbation theory \cite{Bir-Pikus-1974,Voon-Willatzen-2009}. The
resulting Schr\"odinger equation is symbolically the same as the one
in Eq.~(\ref{eq:kp_inf}), but the summation is now restricted to
few bands. The canonical transformation, typically restricted to the second-order
perturbation theory, does not introduce any higher powers of the momentum
operator, so the components of resulting few-band matrix Hamiltonian are still at most quadratic with respect to ${\bf p}$.

Within the envelope function formalism, the interaction of EMF with
charge carriers could be dealt with using two different approaches.
The first approach is to use the envelope functions, obtained from
Eq.~(\ref{eq:kp_inf}), to restore the entire microscopic wavefunction
(\ref{eq:Psi_micr}). Coupling between such wavefunctions due to EMF
could then be evaluated using the microscopic Hamiltonian (\ref{eq:Hem}).
The problem with this approach is that the canonical transformation,
used to decouple far-bands, also transforms the basis (\ref{eq:bwf_basis}),
so that the new basis functions are linear combinations of the old
ones with, in general, time- and coordinate-dependent coefficients.
Under these conditions, the restoration of the microscopic wavefunction
is a tedious task that has to be performed very accurately to preserve
the gauge invariance of the problem.

The other, perhaps more physically transparent approach, is to include
the interaction with the vector and scalar potentials directly into
the envelope function Hamiltonian, Eq.~(\ref{eq:Henv}). This can
always be done directly by starting from the microscopic Hamiltonian
(\ref{eq:Hem}), and then averaging it over the unit cell and performing
the canonical transformation to reduce the number of bands \cite{Bir-Pikus-1974}.
The result, however, can be guessed because the final result - $k\cdot p$
Hamiltonian with external fields - must be gauge-invariant in the
sense introduced in Sec.~\ref{sec:mic_Ham}. More specifically, since
the gauge transformation of the scalar and vector potentials is identical
to unitary transformation of a time-dependent wave function, one can
conclude that the external EMF has to be introduced into $H_{0,nm}({\bf r},{\bf p})$ as
\begin{gather}
H_{0,nm}({\bf x},{\bf p})\rightarrow H_{nm}({\bf x},{\bf p},t)\nonumber\\
=H_{0,nm}\left({\bf x},{\bf p}+\frac{e}{c}{\bf A}({\bf x},t)\right)-e\delta_{nm}\varphi({\bf x},t).\label{eq:subst_em}
\end{gather}
The result is of course very similar to how it was done in the case of the
microscopic Hamiltonian, Eq.~(\ref{eq:Hem}), since the minimal-coupling procedure is a natural
gauge-invariant way to introduce vector and scalar potentials into
a generic Hamiltonian. We wish to emphasize here that once the EMF
is introduced into $H_{0,nm}$, one does not need to restore the microscopic
wavefunctions to perform calculations in the presence of EMF, since
the dynamics of charge carriers in the presence of EMF is fully encoded
by the just obtained envelope function Hamiltonian (\ref{eq:subst_em}).
For example, rates of photoinduced transitions between electronic
states could be obtained by treating terms containing ${\bf A}({\bf r},t)$
and $\varphi({\bf r},t)$ in $H_{nm}({\bf r},{\bf p},t)$ as a perturbation.
More specifically, since components of Hamiltonian $H_{0,nm}$ are at most quadratic with respect to ${\bf p}$, the lowest order operator
of interaction with EMF can be obtained from Eq.~(\ref{eq:subst_em}) as
\begin{gather}
F_{nm}({\bf r},{\bf p},t)=\frac{e}{2c}\left[\frac{\partial H_{0,nm}({\bf r},{\bf p})}{\partial{\bf p}}\cdot{\bf A}({\bf r},t)\right.\nonumber\\
\left.+{\bf A}({\bf r},t)\cdot\frac{\partial H_{0,nm}({\bf r},{\bf p})}{\partial{\bf p}}\right]-e\delta_{nm}\varphi({\bf r},t).\label{eq:eph_coupl}
\end{gather}
For specific gauge choices, Eqs.~(\ref{eq:gauge1}) and (\ref{eq:gauge2}),
this operator becomes
\begin{equation}
F_{1,nm}({\bf r},{\bf p},t)=\frac{e}{c}\left[{\bf A}_{0}\cdot\frac{\partial H_{0,nm}({\bf r},{\bf p})}{\partial{\bf p}}\right]\sin(\omega_{0}t),\label{eq:F1}
\end{equation}
and
\begin{equation}
F_{2,nm}({\bf r},{\bf p},t)=-\delta_{nm}\frac{e\omega_{0}}{c}({\bf r}\cdot{\bf A}_{0})\cos(\omega_{0}t).\label{eq:F2}
\end{equation}
It is a tedious task to directly prove that operators $F_{1}$ and
$F_{2}$, that are first-order in EMF, result in identical
photoinduced transition rates \cite{Aharonov-1979-1553}. However,
since the Hamiltonian is gauge invariant, all physical observables
are exactly gauge independent at EMF of arbitrary strength, and, therefore,
for each order of field strength independently. The leading order
terms of expansion of time-resolved state populations with respect
to the EMF strength do thus produce gauge invariant transition rates.

\section{Two-band Kane model\label{sec:2b_Kane}}

In this section we will explore the general features of how mixing
between electronic bands affects the strength of the electron-EMF
interaction. To this end, we will assume a simplest possible model
with band mixing - a one-dimensional (1D) two-band Kane model, described
by a Hamiltonian \cite{Kane-1956-82,Cercignani-Gabetta-2007,Cragg-2010-313}
\begin{equation}
H_{0}=\left(\frac{E_{g}}{2}+\frac{p^{2}}{2m_{*}}\right)\sigma_{z}+\frac{P}{m}p\sigma_{x},\label{eq:H0_2bKane}
\end{equation}
where $p=-i\hbar\frac{\partial}{\partial x}$ and the Pauli matrices
are
\begin{equation}
\sigma_{x}=\begin{bmatrix}0 & 1\\
1 & 0
\end{bmatrix},\,\sigma_{y}=\begin{bmatrix}0 & -i\\
i & 0
\end{bmatrix},\,\sigma_{z}=\begin{bmatrix}1 & 0\\
0 & -1
\end{bmatrix}.\label{eq:Pauli_mat}
\end{equation}
In Eq.~(\ref{eq:H0_2bKane}), $E_{g}$ is the bulk bandgap energy,
$m_{*}$ is the effective mass of carriers (assumed identical for
electrons and holes), and $P$ is the Kane momentum. The free electron
mass is denoted by $m$. Rescaling length and energy results in a
dimensionless Hamiltonian
\begin{equation}
H_{0}=(1+p^{2})\sigma_{z}+\alpha p\sigma_{x},\label{eq:H0_Kane_unitless}
\end{equation}
where $\alpha$ is the dimensionless Kane momentum and $p=-i\frac{\partial}{\partial x}$.
The operators of the lowest-order electron-EMF interaction, Eqs.~(\ref{eq:F1})
and (\ref{eq:F2}), become, respectively
\begin{equation}
F_{1}(x,p,t)=A_{0}\left[2p\sigma_{z}+\alpha\sigma_{x}\right]\sin(\omega_{0}t),\label{eq:F1_Kane}
\end{equation}
and
\begin{equation}
F_{2}(x,p,t)=-A_{0}x\omega_{0}I_{2}\cos(\omega_{0}t),\label{eq:F2_Kane}
\end{equation}
where $I_{2}$ is the $2\times2$ unit matrix. These expressions were
also made dimensionless by rescaling $A_{0}$.

\subsection{Continuous 1D Kane model\label{sub:Cont_Kane}}

\begin{figure}
\includegraphics[width=8.7cm]{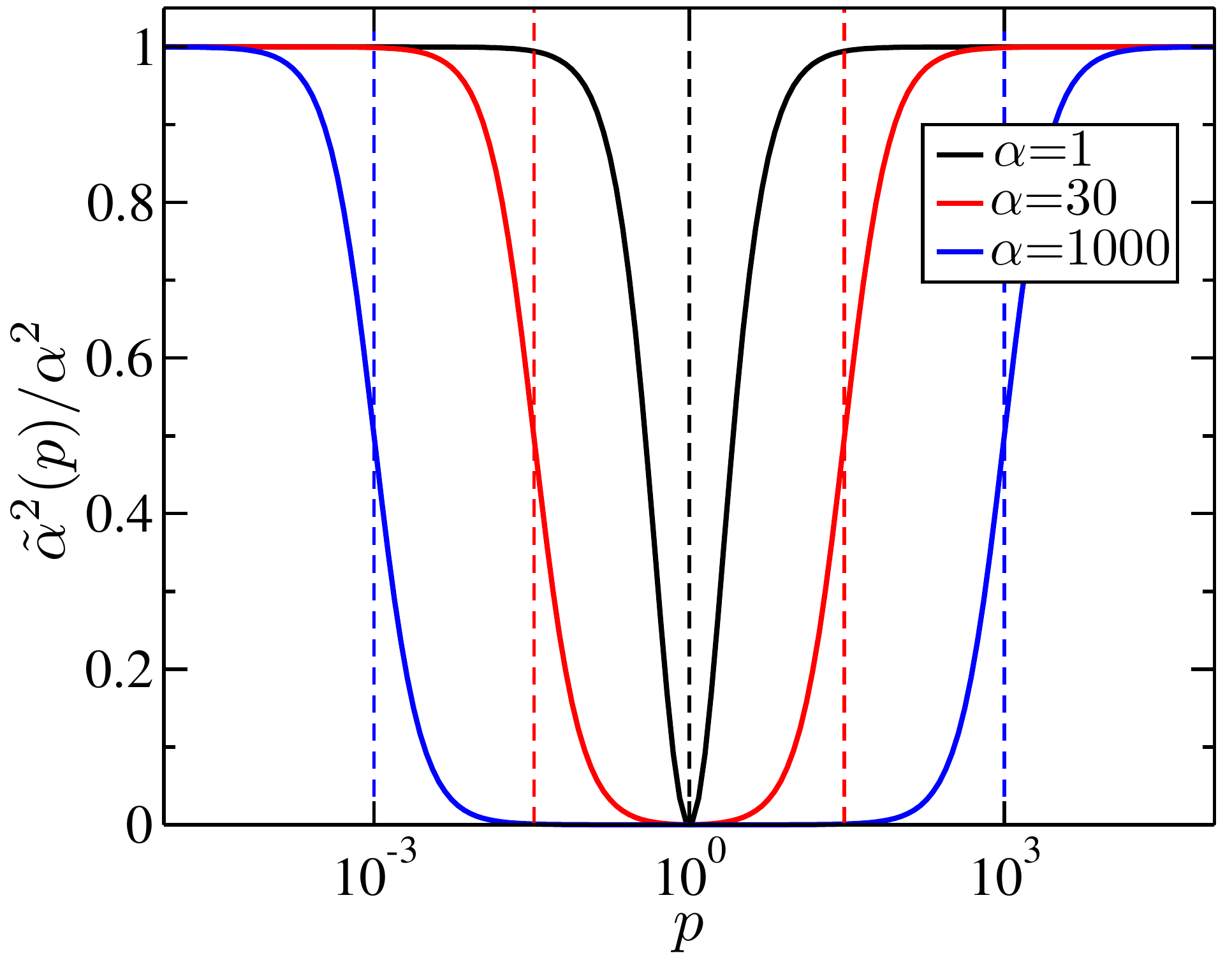}\caption{\label{fig:dp_cont}
Square of the normalized effective Kane momentum as a function of $p$ for three different values of $\alpha=1$, $30$ and $1000$. The vertical dashed lines correspond to $p=\alpha^{-1}$ and $p=\alpha$ for each specific value of $\alpha$.
}
\end{figure}

We first consider the effect of band mixing on the strength of electron-EMF
coupling in the continuous Kane model, i.e., coordinate $x$ is
unrestricted. In this case, the length gauge, Eq.~(\ref{eq:F2_Kane}),
does result in ill-defined transition amplitudes since scalar potential
grows linearly with $x$. We, therefore, choose the velocity gauge, Eq.~(\ref{eq:F1_Kane}).
The amplitude of photoinduced transitions between the eigenstates
of Hamiltonian~(\ref{eq:H0_Kane_unitless}) is then encoded by operator
\begin{equation}
f_{1}=2p\sigma_{z}+\alpha\sigma_{x},\label{eq:f1}
\end{equation}
so that Eq.~(\ref{eq:F1_Kane}) becomes $F_{1}=A_{0}f_{1}\sin(\omega_{0}t)$.
Since the momentum $p$ becomes a parameter in the continuous Kane
model, energies of eigenstates of Hamiltonian~(\ref{eq:H0_Kane_unitless})
are functions of this parameter
\begin{equation}
E_{\pm}(p)=\pm\sqrt{\left(1+p^{2}\right)^{2}+\alpha^{2}p^{2}},\label{eq:Epm_cont}
\end{equation}
and the corresponding wavefunctions (bi-spinors) are
\begin{gather}
\Psi_{\pm}(p)=\left[\alpha^{2}p^{2}+\left(E_{\pm}(p)-1-p^{2}\right)^{2}\right]^{-1/2}\nonumber\\
\times\begin{bmatrix}\alpha p\\
E_{\pm}(p)-1-p^{2}
\end{bmatrix}.\label{eq:eigf_spinor}
\end{gather}
The effective Kane momentum can now be defined via a transition intensity as 
\begin{equation}
\mbox{\ensuremath{\tilde{\alpha}}}^{2}(p)=\left|\langle\Psi_{+}(p)|f_{1}|\Psi_{-}(p)\rangle\right|^{2}=\frac{\alpha^{2}(1-p^{2})^{2}}{(1+p^{2})^{2}+\alpha^{2}p^{2}}.\label{eq:Ip_Kanec}
\end{equation}
Fig.~\ref{fig:dp_cont} shows the square of the normalized effective
Kane momentum as a function of momentum $p$ for three different values
of Kane momentum: $\alpha=1$ (sold black line), $\alpha=30$ (solid
red line) and $\alpha=1000$ (solid blue line). 
%
%
As a function of $p$, $\tilde{\alpha}^{2}(p)$ is seen to have three
distinct regimes: (i) plateau $\tilde{\alpha}^{2}/\alpha^{2}=1$ at
very small momenta, (ii) plateau $\tilde{\alpha}\approx0$ at intermediate
values of $p$, and (iii) $\tilde{\alpha}^{2}/\alpha=1$ at very large
momenta. These regimes are especially pronounced at $\alpha\gg1$.

If momentum $p$ is not too large, i.e., in the first two regimes,
the Hamiltonian (\ref{eq:H0_Kane_unitless}) and the electron-EMF
coupling (\ref{eq:f1}) can be approximated as $H_{0}=\sigma_{z}+\alpha p\sigma_{x}$
and $f_{1}=\alpha\sigma_{x}$, respectively. This approximation is
valid at $p\ll\alpha$. Since $f_{1}$ is proportional to $\sigma_{x}$,
it most efficiently couples spinors that are eigenfunctions of $\sigma_{z}$.
Since $H_{0}\approx\sigma_{z}$ at $p\ll\alpha^{-1}$, the electron-EMF
coupling is most efficient is this regime, resulting in $|\langle\Psi_{+}|f_{1}|\Psi_{-}\rangle|=\alpha$.
We thus associate condition $p\ll\alpha^{-1}$ with the first regime
in Fig.~\ref{fig:dp_cont}, where the Kane momentum is not renormalized,
i.e., $\tilde{\alpha}^{2}=\left|\langle\Psi_{+}|f_{1}|\Psi_{-}\rangle\right|^{2}=\alpha^{2}$.
On the other hand, $H_{0}\approx\alpha p\sigma_{x}$ at $\alpha^{-1}\ll p\ll\alpha$,
so that the eigenstates of $H_{0}$ are not coupled by $f_{1}\propto\sigma_{x}$
in this regime, resulting in $\tilde{\alpha}=0$. We thus associate
condition $\alpha^{-1}\ll p\ll\alpha$ with the second regime in Fig.~\ref{fig:dp_cont}.
Electron-EMF coupling is heavily suppressed in this regime resulting
in $\tilde{\alpha}^{2}/\alpha^{2}\approx0$. In the both first and
second regimes, accounting for neglected terms in $H_{0}$ and $f_{1}$
, e.g., the first term in Eq.~(\ref{eq:f1}), results in negligible
corrections to already obtained values of $\tilde{\alpha}$. 

At first glance, the third regime, $p\gg\alpha$, has to be very similar
to the first one. Indeed, Hamiltonian~(\ref{eq:H0_Kane_unitless})
becomes diagonally dominated again, $H_{0}\approx p^{2}\sigma_{z}$,
so one might follow the considerations above to conclude that $\tilde{\alpha}^{2}/\alpha^{2}=1$
in this regime. This result is indeed correct, although it turns out
that the logic used when discussing the first regime is not quite
applicable here. More specifically, the second term in Eq.~(\ref{eq:f1})
dominated the transition intensity in the first regime. In the third
regime, however, the two terms of $f_{1}$ produce contributions to
the transition intensity that are of the same order of magnitude.
In fact, the contributions from $2p\sigma_{z}$ and $\alpha\sigma_{x}$
to the matrix element in Eq.~(\ref{eq:Ip_Kanec}) are of the opposite
signs, and the contribution of the former term is twice as large in
magnitude as the contribution of latter one. Therefore, the value
of matrix element in Eq.~(\ref{eq:Ip_Kanec}) changes from $+1$
to $-1$ when switching from the first regime to the third one. The
third regime is thus physically rather distinct from the first one,
although the resulting transition intensity is the same since Eq.~(\ref{eq:Ip_Kanec})
is insensitive to the phase of the transition matrix element.

With regards to the third regime, we would like to note here
that even though this regime, $p\gg\alpha$, naturally appears in
Hamiltonian (\ref{eq:H0_2bKane}), we expect it to be of less practical
importance than the first two regimes, at least when treated within
the envelope function approach. Indeed, the Kane momentum $P=\hbar\langle u_{c}|\nabla|u_{v}\rangle$
can be thought of as a certain characteristic momentum associated
with a single unit cell. Then, the condition $p\gg\alpha$ can only
be realized when the wavelength corresponding to the envelope wavefunction is comparable or smaller than the size of a single unit cell. At these conditions, the envelope function formalism
breaks down. As an example, one can show that for the third
regime to be realized in Eq.~(\ref{eq:H0_2bKane}) in case of $\rm{PbSe}$ or $\rm{PbS}$
quantum dots \cite{Kang-1997-1632}, one has to have the quantum confinement
energies significantly exceeding $1$ eV. As is clearly seen from
the bulk dispersion relations \cite{Suzuki-1995-1249,Kanazawa-1998-5997},
the standard $k\cdot p$ approximation for these materials becomes
inaccurate at such energies and higher-energy bands have to be explicitly
added into consideration.

\subsection{Spatially confined 1D Kane model\label{sub:conf_Kane}}

In this subsection we analyze the effect of band mixing on
the renormalization of the strength of electron-EMF interaction in the presence of quantum confinement. To this end, we again adopt the two-band 1D Kane model, Eq.~(\ref{eq:H0_Kane_unitless}).
To introduce the size quantization, a wavefunction is required to
vanish at $x=\pm L/2$, where $L$ is the quantum confinement length.
Quantum-confined levels are then found as follows. First, all the
\emph{continuous} solutions for Hamiltonian (\ref{eq:H0_Kane_unitless})
are found for a given energy. Since the resulting characteristic equation
is of the forth order with respect to momentum $p$, four linearly
independent continuous solutions, each being a bi-spinor, are obtained.
Second, a linear combination of these solutions is required to vanish
at $x=\pm L/2$, which yields a homogeneous system of linear equations
with four unknowns. The resulting secular equation yields the discrete
energy spectrum, corresponding to the levels of the quantum-confined
two-band 1D Kane model. This procedure is described in more detail in \ref{app:Kane_conf}.

\begin{figure}
\includegraphics[width=8.7cm]{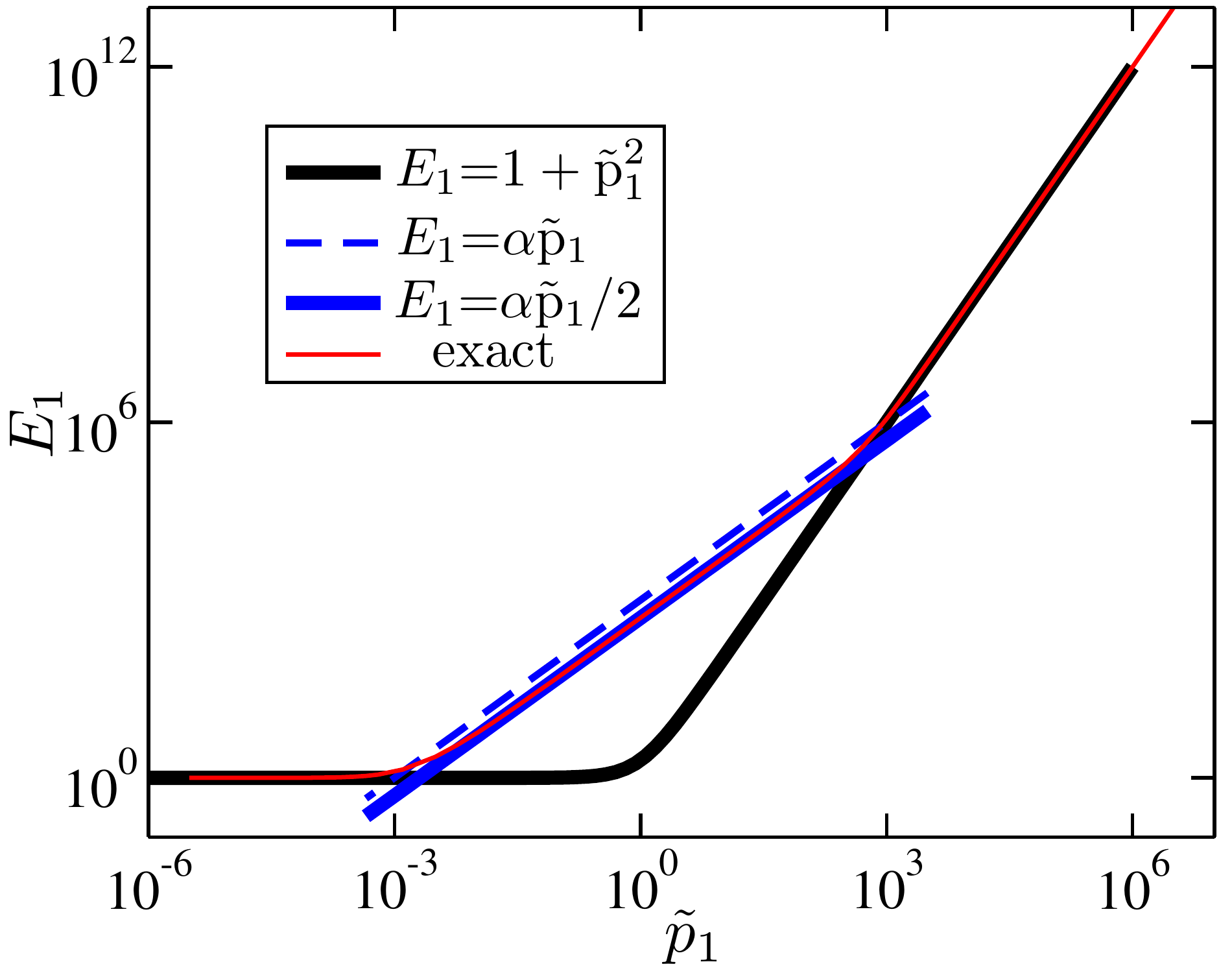}
\caption{\label{fig:energy_kane_conf}
Energy of the lowest positive-energy state in a spatially confined Kane model ($\alpha=1000$), plotted as a function of the effective momentum, $\tilde{p}_1=\pi/L$. Thin red line is the exact numerical result. Thick black line is the expected asymptotic behavior in the the regimes of the weak and strong confinement, $E_1=1+\tilde{p}_1^2$. Dashed blue lines represents the ``naive" spatial quantization in the intermediate regime, $E_1\rm{=}\alpha\tilde{p}_1$. Thick blue lines shows the correct asymptotic dependence in the intermediate regime, $E_1\rm{=}\alpha\tilde{p}_1/2$.
}
\end{figure}

The dependence of the energy of the lowest conduction band state on effective momentum $\tilde{p}_1=\pi/L$ is shown in Fig.~\ref{fig:energy_kane_conf}. 
%
%
If the band mixing is absent (i.e., $\alpha=0$) or could be neglected,
then Hamiltonian (\ref{eq:H0_Kane_unitless}) encodes two uncoupled
bands, each having simple parabolic dependence of energy on momentum.
At these conditions, the problem of size quantization reduces to the
textbook ``particle in a box'' problem with the result $E_{\pm n}=\pm(1+\tilde{p}_{n}^{2})$,
where $n=1,2,3,...$ and $\tilde{p}_{n}=\pi n/L$. Hamiltonian of the continuous Kane model (\ref{eq:H0_Kane_unitless})
is diagonally dominated at either $p\ll\alpha^{-1}$ or $p\gg\alpha$,
so, expectedly, the exact result for $E_{1}$(thin red line) is well
described by $E_{1}=(1+\tilde{p}_{1}^{2})$ at very small or very
large effective momenta.

The just obtained approximate expression for the energy at very large or very small momenta could be obtained from Eq.~(\ref{eq:Epm_cont}) by simply substituting $p\rightarrow\tilde{p}_1$ and then assuming small $\alpha$ limit. Analogously, one would expect that Eq.~(\ref{eq:Epm_cont}) yields $E_{1}\approx\alpha\tilde{p}_{1}$ in the intermediate confinement
regime, $\alpha^{-1}\ll\tilde{p}_{1}\ll\alpha$. However, this dependence,
shown by dashed blue line in Fig.~(\ref{fig:energy_kane_conf}) clearly
deviates from the exact result (red line) by a constant prefactor. That the behavior
of the system in this regime is more involved can already be suspected
from the following consideration. In the intermediate regime, Hamiltonian
(\ref{eq:H0_Kane_unitless}) could be approximated as $H_{0}\approx\alpha p\sigma_{x}$.
However, the quantum-confined states for such a Hamiltonian are not
well-defined because this Hamiltonian, being only first-order with
respect to momentum, yields only two linearly independent continuous
solutions for each energy, and so the boundary conditions can not be satisfied.
Therefore, the diagonal terms of the Hamiltonian (\ref{eq:H0_Kane_unitless})
cannot be simply dropped even if $\alpha^{-1}\ll\tilde{p}_{1}\ll\alpha$.
This leads to a suspicion that one cannot just take energies $E(p)=\pm\alpha p$
corresponding to continuous solutions of $H_{0}\approx\alpha p\sigma_{x}$,
and then ``quantum-confine'' them by setting $p=\tilde{p}_{n}$.
An accurate analysis of the effect of quantum confinement in this
regime, provided in \ref{app:Kane_conf}, results in $E_{1}=\alpha\tilde{p}_{1}/2$.
This result is plotted by a solid blue line in Fig.~\ref{fig:energy_kane_conf},
demonstrating an excellent agreement with the exact numerical result (thin red line). 

\begin{figure}
\includegraphics[width=8.7cm]{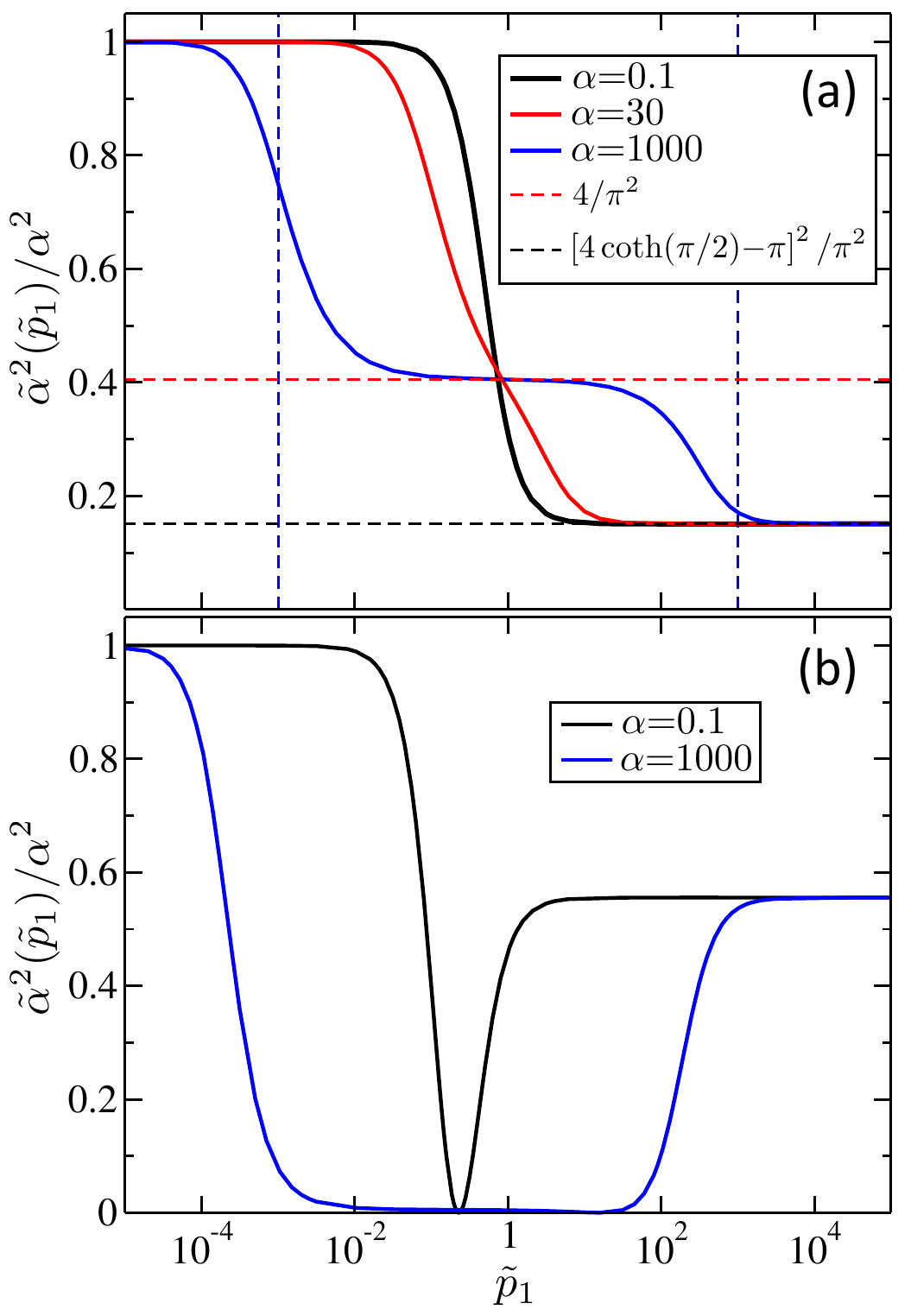}\caption{\label{fig:dp_conf}
Panel (a) shows the normalized transition intensity for the lowest-energy transition, $E_{-1}\rightarrow E_{1}$, as a function of $\tilde{p}_{1}=\pi/L$ for three different values of Kane momentum: $\alpha=0.1$, $10$ and $1000$. The vertical dashed lines correspond to $\tilde{p}_{1}=\alpha^{-1}$ and $\tilde{p}_{1}=\alpha$ for $\alpha=1000$. Panel (b) is the same except that it is not for the lowest transition, but for the $E_{-5}\rightarrow E_{5}$ transition.
}
\end{figure}

Fig.~\ref{fig:dp_conf}(a) shows the normalized transition intensity
for the lowest energy transition, $E_{-1}\rightarrow E_{1}$, as a
function of the effective momentum $\tilde{p}_{1}$.
%
%
In the most illustrative case, $\alpha\gg1$, one can again see three distinct regimes corresponding to those already
observed in Fig.~\ref{fig:energy_kane_conf}. The general trend of
suppressing transition intensity when going from the weak ($\tilde{p}_{1}\ll\alpha^{-1}$)
to intermediate ($\alpha^{-1}\ll\tilde{p}_{1}\ll\alpha$) quantum
confinement is qualitatively similar to that observed for the continuous
Kane model in Sec.~\ref{sub:Cont_Kane}. Quantitatively, however, the
continuous and spatially confined Kane models are different in the
intermediate regime. As discussed above, even though Hamiltonian (\ref{eq:H0_Kane_unitless})
is off-diagonally dominated in this regime, the diagonal terms cannot
be completely disregarded in the spatially confined case. Accounting
for these terms results in only a partial suppression of the transition
intensity in the intermediate regime. More quantitatively, the effective
Kane momentum becomes $\tilde{\alpha}^{2}/\alpha=4/\pi^{2}$ in the
limit $\alpha\gg1$, as derived in \ref{app:Kane_conf}. This
asymptotic result, plotted by a horizontal dashed red line in Fig.~\ref{fig:dp_conf}(a),
is seen to be in agreement with the exact dependence of $\tilde{\alpha}^{2}/\alpha^{2}$
on $\tilde{p}_{1}$ (blue line) in the intermediate confinement regime.

The transition intensity for the lowest transition drops even further
when switching from the intermediate to the strong ($\tilde{p}_{1}\gg\alpha$)
confinement regime. Asymptotically, $\tilde{\alpha}^{2}/\alpha^{2}=\left[4\coth(\pi/2)-\pi\right]^{2}/\pi^{2}\approx0.15$
in this regime when $\alpha\gg1$, as derived in \ref{app:Kane_conf}
and shown as a horizontal dashed black line in Fig.~\ref{fig:dp_conf}(a).
This is rather different from $\tilde{\alpha}^{2}/\alpha^{2}=1$ obtained
for the continuous Kane model. The discrepancy can ultimately be traced
back to the fact that the band-mixing term in Eq.~(\ref{eq:H0_Kane_unitless}),
i.e., $\alpha p\sigma_{x}$, does of course conserve momentum in the
continuous case, but can couple states of different \emph{effective}
momenta $\tilde{p}_{n}$ in the spatially confined case. This latter
effect has to be less important for higher-energy transitions where
the effective momentum becomes more and more ``conserved''. One,
therefore, can expect that the difference between continuous and spatially-confined
Kane models with regards to the renormalization of transition intensity
will become less pronounced at large transition energies. Fig.~\ref{fig:dp_conf}(b)
shows the normalized intensity for the $E_{-5}\rightarrow E_{5}$
transition, and this is clearly seen that the dependence of $\tilde{\alpha}^{2}/\alpha^{2}$
on $\tilde{p}_1$ becomes very reminiscent to the results obtained previously
for the continuous Kane model, Fig.~\ref{fig:dp_cont}. The spatially
confined and continuous Kane models can therefore yield similar renormalization
of the transition intensities at large transition energies. The lowest
transition, however, experiences the most effect of the quantum confinement,
resulting in a renormalization that is quite different for the two
models considered in this section.

\section{Lead chalcogenide quantum dots\label{sec:lead_chalc}}

In this section we consider a more realistic and experimentally relevant
example - the effect of band mixing on the strength of the electron-EMF
coupling in lead selenide ($\rm{PbSe}$) and lead sulfide ($\rm{PbS}$) spherically symmetric SNs (i.e., quantum
dots). A four-band $k\cdot p$ Hamiltonian to describe electronic excitations
in bulk lead chalcogenide ($\rm{PbSe}$, $\rm{PbS}$ and $\rm{PbTe}$) was originally introduced
by Dimmock \cite{Dimmock-1964-A821,Dimmock-1971}, and Mitchell and
Wallis \cite{Mitchell-1966-581}. The spherically symmetric version
of this Hamiltonian reads as
\begin{equation}
H_{0}=\begin{bmatrix}\left(\frac{E_{g}}{2}+\frac{p^{2}}{2m_{c}}\right)I_{2} & \frac{P}{m}(\boldsymbol{\sigma}\cdot{\bf p})\\
\frac{P}{m}(\boldsymbol{\sigma}\cdot{\bf p}) & -\left(\frac{E_{g}}{2}+\frac{p^{2}}{2m_{v}}\right)I_{2}
\end{bmatrix},\label{eq:Ham_KW}
\end{equation}
where $m_{c}$ and $m_{v}$ are far-band contributions to the electron
and hole effective masses, respectively. Bulk bandgap energy is denoted
by $E_{g}$. Momentum operator is ${\bf p}=-i\hbar\nabla$. Kane momentum
is denoted by $P$. Pauli vector is defined as $\boldsymbol{\sigma}=[\sigma_{x},\sigma_{y},\sigma_{z}]$,
and $I_{2}$, as defined before, is the $2\times2$ unit matrix. Energies,
$E_{n}$, and wavefunctions, $\Psi_{n}(r)$, of the quantum-confined
levels for a spherically symmetric lead chalcogenide SNs (e.g., quantum
dots) can be found by applying boundary conditions $\left.\Psi(r)\right|_{|r|=a}=0$,
where $a$ is the quantum dot radius. The entire procedure of finding
such quantum-confined energies and wavefunctions is described in great
detail in the seminal paper by Kang and Wise \cite{Kang-1997-1632}, and briefly outlined
in \ref{sec:Kang-Wise}, putting particular emphasis on the symmetry considerations.

Once the wavefunctions (four-spinors) and the corresponding energies
are obtained for the quantum-confined levels within the conduction
($E>0$) and the valence ($E<0$) bands, the electron-EMF coupling
can be calculated. Operator for the lowest-order electron-EMF interaction,
Eq.~(\ref{eq:eph_coupl}), becomes
\begin{gather}
F({\bf r},{\bf p},t)=\frac{e}{c}\begin{bmatrix}\frac{{\bf p}\cdot{\bf A}({\bf r},t)+{\bf A}({\bf r},t)\cdot{\bf p}}{2m_{c}}I_{2} & \frac{P}{m}\left(\boldsymbol{\sigma}\cdot{\bf A}({\bf r},t)\right)\\
\frac{P}{m}\left(\boldsymbol{\sigma}\cdot{\bf A}({\bf r},t)\right) & -\frac{{\bf p}\cdot{\bf A}({\bf r},t)+{\bf A}({\bf r},t)\cdot{\bf p}}{2m_{v}}I_2
\end{bmatrix}\nonumber\\
-e\varphi({\bf r},t)I_{4},
\end{gather}
where $I_{4}$ is the $4\times4$ unit matrix. Vector and scalar potentials in velocity and length gauges are given by Eqs.~(\ref{eq:gauge1}) and (\ref{eq:gauge2}), respectively. For definiteness we assume ${\bf A}_{0}=A_{0}\hat{{\bf z}}$, where
$\hat{{\bf z}}$ is the unit vector along z-axis. Then, amplitudes
of photoinduced transitions between the eigenstates of Hamiltonian~(\ref{eq:Ham_KW})
are then encoded by operators
\begin{equation}
f_{1}({\bf r},{\bf p})=\frac{e}{c}\begin{bmatrix}\frac{p_{z}}{m_{c}}I_{2} & \frac{P}{m}\sigma_{z}\\
\frac{P}{m}\sigma_{z} & -\frac{p_{z}}{m_{v}}I_2
\end{bmatrix},\label{eq:f1_KW}
\end{equation}
and
\begin{equation}
f_{2}({\bf r},{\bf p})=ez\frac{\omega_{0}}{c}I_{4},\label{eq:f2_KW}
\end{equation}
so that $F=A_{0}f_{1}\sin(\omega_{0}t)$ and $F=A_{0}f_{2}\cos(\omega_{0}t)$
for the velocity and length gauge, respectively.

In what follows we will focus on the lowest-energy photoinduced transition,
which is most relevant for photoluminescence. This is the transition
between the highest energy quantum-confined electronic state in the
valence band and the lowest energy quantum-confined state in the conduction
band. These states are designated as $(n,J,J_{z},\pi)_{v}=(1,1/2,\pm1/2,1)_{v}$
and $(n,J,J_{z},\pi)_{c}=(1,1/2,\pm1/2,-1)_{c}$, respectively, where
$n$ is the radial quantum number, $J$ and $J_{z}$ are the total
angular momentum and its projection onto z-axis, respectively, and $\pi$ is parity
\cite{Kang-1997-1632}. The respective wavefunctions are $\Psi_{-1}$
and $\Psi_{1}$. Each of these states is doubly degenerate with respect
to the projection of the total angular momentum. However, this projection
is conserved during the transition since we chose ${\bf A}_{0}=A_{0}\hat{{\bf z}}$.
Henceforth, we consider the photoinduced transition between states
with $J_{z}=+1/2$ for definiteness. As previously, we can define
an effective energy-dependent Kane momentum as (substituting $\hbar\omega_{0}$
with transition energy $E$)
\begin{equation}
\tilde{P}_{i}(E)=\frac{cm}{e}\langle\Psi_{1}|f_{i}|\Psi_{-1}\rangle,\,\,i=1,2.\label{eq:Pi}
\end{equation}

As discussed in Sec.~\ref{sec:envelope}, the choice of gauge does
not affect the transition intensities. We tested numerically that
this is indeed the case for the two gauge choices, Eqs.~(\ref{eq:f1_KW})
and (\ref{eq:f2_KW}), up to numerical round-off errors, so that
\begin{equation}
\tilde{P}^{2}(E)=\tilde{P}_{1}^{2}(E)=\tilde{P}_{2}^{2}(E).\label{eq:PPP}
\end{equation}
The normalized transition intensity, $\tilde{P}^{2}(E)/P^{2}$, is
plotted in Fig.~\ref{fig:P_E} for $\rm{PbSe}$ and $\rm{PbS}$ with material parameters
taken from Ref.~\cite{Kang-1997-1632}. The observed behavior -- transition
intensity decreases with increasing degree of quantum confinement
-- is qualitatively similar to that observed for the two-band Kane model
in Sec.~\ref{sub:conf_Kane}. In particular, making Hamiltonian (\ref{eq:Ham_KW})
dimensionless similarly to how it was done in the beginning of Sec.~\ref{sec:2b_Kane},
one obtains the dimensionless Kane momentum matrix element as $\alpha=2P\sqrt{\tfrac{4m_{*}}{m^{2}E_{g}}}$,
where $m_{*}$ is the characteristic far-band contribution to carrier
effective masses. Substituting numerical values for $\rm{PbSe}$ and $\rm{PbS}$ parameters
to this expression one obtains $\alpha\approx1-2$. The intermediate
regime of Kane momentum renormalization does not appear for such small
values of $\alpha$ in the spatially confined Kane model, see red
and black curves in Fig.~\ref{fig:dp_conf}(a). Based on this, one
would expect a simple monotonic featureless decrease of $\tilde{P}^{2}(E)/P^{2}$
as a function of $E$ for lead chalcogenide QDs, starting from $\tilde{P}^{2}(E)=P^{2}$
at low energies and converging to some finite value $0<\tilde{P}^{2}(E)/P^{2}<1$
at $E\rightarrow\infty$. This is exactly what is observed in Fig.~\ref{fig:P_E}.

Finally, we would like to discuss the dependence of the renormalized
Kane momentum and the renormalized transition dipole moment on energy.
Combining Eqs.~(\ref{eq:f2_KW}) and (\ref{eq:Pi}) with the direct
consequence of gauge invariance, Eq.~(\ref{eq:PPP}), one obtains
\begin{equation}
\tilde{P}^{2}(E)\propto E^{2}\tilde{d}^2(E),\label{eq:PEd}
\end{equation}
where $\tilde{d}(E)=\left|\langle\Psi_{1}|ez|\Psi_{-1}\rangle\right|$ is the
dipole moment corresponding to the lowest-energy transition. Eq.~(\ref{eq:PEd})
is exactly the relation between the effective transition dipole moment
and the effective Kane momentum we guessed in Introduction. In
particular, this relation guarantees that Eqs.~(\ref{eq:kr_P}) and
(\ref{eq:kr_d}) produce exactly the same spontaneous emission rates.
On the other hand, since $\tilde{P}(E)$ already decays with $E$, as is seen in Fig.~\ref{fig:P_E}, Eq.~(\ref{eq:PEd}) implies that $\tilde{d}(E)$ decays \emph{rapidly}
with $E$. This suggests that the renormalized dipole moment is not
especially convenient to use when describing the strength of electron-EMF
interaction in narrow-gap SNs. Effective Kane momentum $\tilde{P}(E)$ depends on
$E$ much less strongly and, perhaps, could even be approximated by a constant
within an experimentally relevant range of energies.

\section{Discussion\label{sec:Discussion}}

In this section we will discuss two example approaches of calculation of the electron-EMF coupling strength, which ignore the effect of band mixing, and clarify exactly why they produce inaccurate results.

\subsection{Velocity gauge\label{sub:velocity_gauge}}

\begin{figure}
\includegraphics[width=8.7cm]{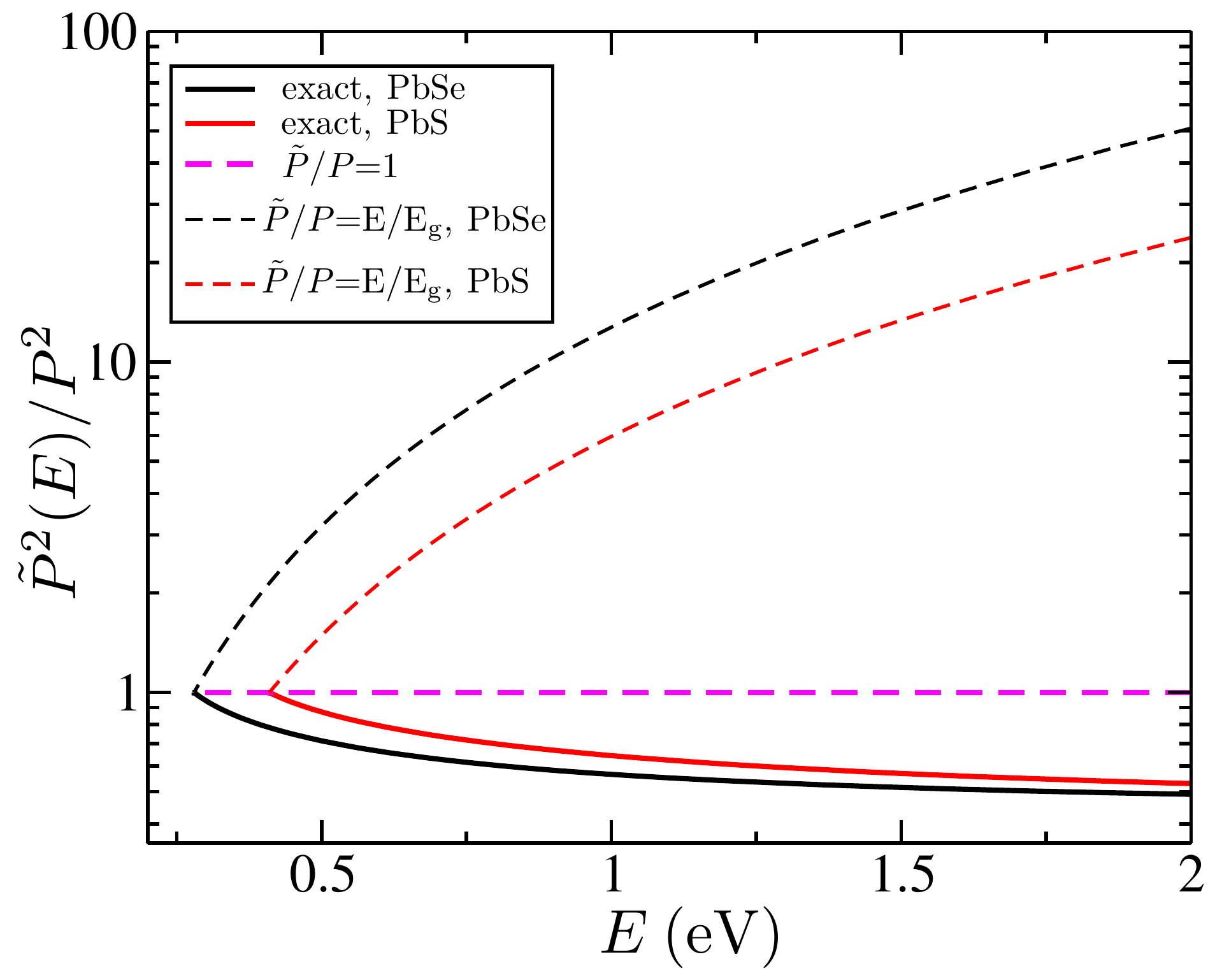}\caption{\label{fig:KW_compar}
Energy-dependent normalized transition intensity for the lowest optical transition in $\rm{PbSe}$ and $\rm{PbS}$ QDs. Exact numerical results are given by solid lines. Dashed magenta line corresponds to neglecting band mixing in the velocity gauge. Dashed black and red lines correspond to neglecting band mixing in the length gauge.
}
\end{figure}

The first approach is based on the velocity gauge, Eq.~(\ref{eq:gauge1}), where the scalar
potential vanishes exactly. The typical calculation of coupling strength
in this gauge, which neglects band mixing, proceeds as follows. Kane
momentum matrix element in Hamiltonian (\ref{eq:Ham_KW}) is set to
zero, thus neglecting band mixing. At these conditions, finding size-quantized
levels becomes very simple since the entire problem reduces to a single-band
effective mass approximation. Specifically, the wavefunctions corresponding
to the lowest conduction band and the highest valence band states
with total angular momentum projection of $J_{z}=+1/2$ are
\begin{gather}
\Psi_{1}\propto[j_{0}(\kappa r),0,0,0]^{T},\nonumber \\
\Psi_{-1}\propto[0,0,j_{0}(\kappa r),0]^{T},\label{eq:Psi_no_mix}
\end{gather}
 respectively, where $\kappa=\pi/a$ and $j_0(x)=\sin x/x$ is the zeroth-order spherical Bessel function of the first kind. These wavefunctions can now
be substituted into Eq.~(\ref{eq:Pi}) with $f_{i}=f_{1}$ to evaluate
the coupling between them due to interaction with EMF. The result,
$\tilde{P}(E)\equiv P$, is shown as a horizontal dashed magenta line
in Fig.~\ref{fig:KW_compar}. 
%
%
Neglecting the band mixing in the velocity gauge does thus result
in the absence of the Kane momentum renormalization. Exact results,
obtained in the previous section for $\rm{PbSe}$ and $\rm{PbS}$ are shown by solid
black and solid red lines, respectively. Comparing these approximate
and exact results one can conclude that the effect of band mixing
on the strength of the electron-EMF coupling can not in general be
disregarded except for very large QDs where the energy of the lowest optical
transition approaches the bulk bandgap energy $E_{g}$. However, since the
renormalized $\tilde{P}(E)$ does not deviate from $P$ by more than
a factor of $\sim$2 for $\rm{PbSe}$ and $\rm{PbS}$, one might be satisfied with
assuming $\tilde{P}(E)\approx P$ depending on the required accuracy.

However, what is not immediately clear at the first glance is that
the considered approach is also not gauge-invariant. As discussed
in Sec.~\ref{sec:envelope}, the gauge invariant formulation could
be obtained from Hamiltonian (\ref{eq:Ham_KW}) via substitution ${\bf p}\rightarrow{\bf p}+\frac{e}{c}{\bf A}({\bf r},t)$
and adding the scalar potential. The resulting Hamiltonian is the
polynomial of the second order with respect to the vector potential.
To arrive at the approximate approach just described above, one needs to
set $P$ to zero, but only in the polynomial terms that are of zeroth-order
with respect to ${\bf A}({\bf r},t)$, which represent the material
part of the Hamiltonian. Obviously this breaks the gauge invariance
since the resulting approximate Hamiltonian cannot be reduced to one
where the vector potential only enters via ${\bf p}+\frac{e}{c}{\bf A}({\bf r},t)$.
To illustrate by an example the danger of breaking the gauge invariance, we start from the velocity gauge, Eq.~(\ref{eq:gauge1}),
and perform the gauge transformation, Eq.~(\ref{eq:gauge_trans}),
with $f({\bf r},t)=A_{1}z\sin(\omega_{0}t)$ where $A_{1}$ is an
arbitrary value. The resulting vector and scalar potentials are ${\bf A}({\bf r},t)=(A_{0}+A_{1})\hat{{\bf z}}\sin(\omega_{0}t)$
and $\varphi({\bf r},t)=-\frac{\omega_{0}}{c}z\cos(\omega_{0}t)$.
The scalar potential does not couple wavefunctions $\Psi_{1}$ and
$\Psi_{-1}$ given by Eq.~(\ref{eq:Psi_no_mix}) since it always enters
the Hamiltonian matrix strictly diagonally, see Eq.~(\ref{eq:subst_em}).
The amplitude of the vector potential upon the gauge transformation,
$A_{0}+A_{1}$, is arbitrary and therefore the transition intensity
could be made as large as desired. Therefore, that the velocity gauge-based
approach described in this subsection does yield results that are
different from the exact ones by no more than a factor of $\sim$2
is solely due to the fortunate choice of the gauge. On the other hand,
the inclusion of the band mixing would allow the scalar potential
to couple $\Psi_{1}$ and $\Psi_{-1}$, resulting in a matrix element
opposite in sign to that produced by the vector potential, thus
compensating the arbitrary increase in the amplitude of the vector potential.

\subsection{Length gauge \label{sec:LengthGauge}}

Another approach to evaluate the strength of the electron-EMF coupling is based on the length gauge, Eq.~(\ref{eq:gauge2}). Calculation of the coupling
strength in this gauge could proceed as follows. The full microscopic
wavefunctions are written as $\Psi_{1}({\bf r})\propto j_{0}(\kappa r)\chi_{ck}({\bf r})$
and $\Psi_{-1}({\bf r})\propto j_{0}(\kappa r)\chi_{vk}({\bf r})$,
i.e., combining Eqs.~(\ref{eq:Psi_no_mix}) and (\ref{eq:Psi_micr}). Coupling
between these two wavefunctions is then given by
\begin{equation}
\langle\Psi_{1}|f_{1}|\Psi_{-1}\rangle=\frac{e\omega_{0}}{c}\int d{\bf r}\,\Psi_{1}^{*}({\bf r})z\Psi_{-1}({\bf r}).
\end{equation}
This integral can be approximately evaluated assuming that the envelope
functions (i.e., spherical Bessel functions) change much more slowly with ${\bf r}$ than the Bloch functions
$\chi_{c(v)k}({\bf r})$. At these conditions, Eq.~(\ref{eq:Pi})
becomes
\begin{equation}
\tilde{P}(E)=\frac{mE}{e\hbar}d,
\end{equation}
where $E$ is the transition energy and
\begin{equation}
d=e\left|\int_{uc}d{\bf r}\,\chi_{ck}^{*}({\bf r})z\chi_{vk}({\bf r})\right|,
\end{equation}
is the transition dipole moment of a single unit cell, corresponding
to the transition between the valence and the conduction bands. Taking
into account the relation $d=\frac{e\hbar P}{mE_{g}}$ \cite{Rosencher-Vinter-2002,Haug-Koch-2009}, one
obtains
\begin{equation}
\tilde{P}^{2}(E)/P^{2}=(E/E_{g})^{2},
\end{equation}
that is the effective Kane momentum grows rapidly with energy, as
is shown by dashed black and red lines for $\rm{PbSe}$ and $\rm{PbS}$, respectively,
in Fig.~\ref{fig:KW_compar}. The obtained results are generally
very different from the exact ones except when $E\approx E_{g}$.
The stark disagreement with the exact results can be understood from
the following considerations. As it was discussed in Sec.~\ref{sec:envelope},
the band mixing is present in Hamiltonian (\ref{eq:Ham_KW}) not only
explicitly via the off-diagonal terms proportional to $P$, but also
implicitly via $m_{c}$ and $m_{v}$, which come about from mixing
with far-bands. Since this mixing is present, it is not correct anymore
to assume that an envelope wavefunction could be converted to its microscopic
representation using Eq.~(\ref{eq:Psi_micr}), as discussed in Sec.~\ref{sec:envelope}.
To clarify this, we consider an example where the contribution of
far-bands is neglected whatsoever. Then, one has $m_{c}=m$ and $m_{v}=-m$
in Eq.~(\ref{eq:Ham_KW}). In the absence of band mixing between the valence and conduction bands (via $P$), the quantum confinement would just shift all the levels by the same amount of
energy, so $E\equiv E_{g}$. At these conditions, $\tilde{P}(E)\equiv P$,
i.e., identical to the result obtained for the velocity gauge.

To conclude this section, we would like to reiterate that there is a single gauge-invariant approach to account for the electron-EMF interaction within the envelope function approximation. This approach consists of transforming Hamiltonian (\ref{eq:Henv}) into Hamiltonian (\ref{eq:subst_em}) via the minimal-coupling procedure. Other approaches, if not reducible to Eq.~(\ref{eq:subst_em}), break the gauge invariance and, therefore, produce unphysically gauge-dependent experimentally measurable quantities such as photoinduced transition rates. This section gives two most typical examples of such approaches where gauge invariance is broken. Other examples could be found in literature. For example, a mixture of the two provided examples was used in Ref.~\cite{Kang-1997-1632}. Specifically, the full microscopic wavefunction was ``restored" by disregarding the mixing with far-bands, which broke the gauge invariance. The transition intensity was then evaluated within the velocity gauge, producing Eq.~(30) in Ref.~\cite{Kang-1997-1632}. That the obtained expression is incorrect can already be recognized by observing that it does not include far-band contributions to effective masses as parameters. The correct coupling operator, Eq.~(\ref{eq:f1_KW}), was obtained from the gauge-invariant Eq.~(\ref{eq:eph_coupl}), and is seen to contain those far-band contributions to effective masses.

\section{Conclusion\label{sec:Conclusion}}

In this work, we have investigated the effect of band mixing within
the envelope function approximation on the strength of the electron-EMF interaction in SNs. Often, the strength of this interaction is considered
to be given by the Kane momentum matrix element, which is
the property of the bulk semiconductor and, therefore, is independent
of size-quantization effects in e.g., semiconductor quantum dots.
In this work we demonstrate that even though such an approximation
could be relatively accurate in wide-bandgap semiconductors (e.g., $\rm{CdSe}$,
$\rm{CdS}$), it breaks down in narrow-gap semiconductors (e.g., $\rm{PbSe}$, $\rm{PbS}$). In particular, we observe that neglecting the band mixing effects can lead
to overestimation of photoinduced transition or spontaneous emission
rates by a factor of $\sim$2 for $\rm{PbSe}$ and $\rm{PbS}$ quantum dots.

To obtain insight into exactly how band mixing affects the strength
of electron-EMF coupling, we first analyzed a simple possible model
with band mixing - two-band 1D Kane model. The effect could
be seen most transparently in the continuous Kane model, i.e., where
no size quantization was present. On a very qualitative approximate
level, one can think that both (i) band mixing terms of the envelope function
Hamiltonian and (ii) the external EMF do perform the same function - they
couple electronic states from different bands (e.g., valence and conduction).
Then, when band mixing terms are weak, e.g., when energies are very close
to band edges, EMF mixes bands most efficiently resulting in the strong
electron-EMF coupling. On the other hand, when the effect of mode
mixing terms is significant (i.e., higher energies), the conduction and
valence states become strongly mixed within the envelope function
formalism. Under these conditions, the external EMF cannot mix such
states any further since they are already strongly mixed into symmetric
and antisymmetric linear combinations of conduction and valence band
states. This results in a diminished strength of the electron-EMF
coupling. The picture becomes much more involved once the spatial
quantization is present, since a single state from a band can
now couple to multiple states within the other band. Nevertheless,
the introduced qualitative picture holds if we substitute energies
of the continuous electronic states with discrete quantum confinement
energies in SNs. Specifically, the electron-EMF coupling is strongest
in large SNs, which corresponds to small confinement energies, and
becomes weaker in smaller SNs where confinement energies become larger.

It turns out that even though the phenomenon of renormalization of electron-EMF interaction by band mixing could be rather involved
physically, it can nevertheless be quantified by a single energy-dependent
function - an effective energy-dependent Kane momentum, $\tilde{P}(E)$.
Such a function in general depends on the material parameters (e.g.,
bulk band gap, effective masses) and on the SN shape. Fig.~\ref{fig:P_E}
shows our numerical results for the effective energy-dependent Kane
momenta for $\rm{PbSe}$ and $\rm{PbS}$ spherically symmetric SNs (i.e., quantum
dots). This renormalized Kane momenta can now be used in all the formulae
previously developed to treat the electron-EMF interaction in semiconductor
quantum dots in the absence of band mixing. For instance, the familiar
expression for rate of spontaneous emission, Eq.~(\ref{eq:kr_P}),
can be modified to properly account for the band mixing by a simple
substitution $P\rightarrow\tilde{P}(E)$, where $\tilde{P}(E)$ for
$\rm{PbSe}/$\rm{PbSe} QDs can be taken from Fig.~\ref{fig:P_E}. Other quantities
related to the electron-EMF coupling strength, e.g., cross sections
of absorption and stimulated emission, are linearly related to the
rate of spontaneous emission by virtue of the Einstein coefficients.
Therefore, the conventional bulk-like expressions for such quantities
could also be modified to account for band mixing via the same substitution.

Finally, we would like to emphasize that a consistent treatment of the electron-EMF coupling must necessarily preserve the gauge invariance. For example, the apparent contradiction between Eqs.~(\ref{eq:kr_P}) and (\ref{eq:kr_d}) originates from approximations that break the gauge invariance of the $k\cdot p$ model. Two typical examples of calculations of the electron-EMF coupling where the gauge invariance is broken are presented and discussed in Sec.~\ref{sec:Discussion}. The requirement of the gauge invariance is not unique to the $k\cdot p$ model, as it is also very important in tight-binding \cite{Boykin-2001-245314,Foreman-2002-165212} and pseudopotential-based calculations \cite{Pickard-2003-196401,Schwerdtfeger-2012-014107}.

\section*{Acknowledgements}

We are grateful to Nikolay Makarov and Oleksandr Isaienko for useful discussion. K.A.V. was supported by the Center for Advanced Solar Photophysics (CASP), an Energy Frontier Research Center funded by the Office of Basic Energy Sciences, Office of Science, US Department of Energy (DOE).

\appendix

\section{Spatially confined 1D Kane model\label{app:Kane_conf}}

The Hamiltonian of the 1D Kane model in the absence of the EM fields
is given by Eq.~(\ref{eq:H0_Kane_unitless}). The spatial confinement
is introduced by requiring the wavefunction to vanish at the ``surface'',
i.e., $\Psi(x=\pm L/2)=0$, where $L$ is the size of the 1D ``quantum
dot''. The general approach to finding the eigenstates of such a
problem is as follows. First, bulk solutions (i.e., without boundary
conditions) are found for a given energy $E$. Since the Hamiltonian
is a $2\times2$ matrix quadratic with respect to momentum $p$, there
is four bulk solutions, each represented by a two-component spinor.
Specifically, there are two propagating and two evanescent plane waves
at $|E|>1$ , whereas all the waves are evanescent at $|E|<1$. In
what follows, we will only consider the former case. Characteristic
equation for Hamiltonian (\ref{eq:H0_Kane_unitless}) is

\begin{equation}
E^{2}=(1+p^{2})^{2}+\alpha^{2}p^{2}.\label{eq:bulk_disp}
\end{equation}
Solving this equation with respect to momentum produces
\begin{equation}
p^{2}=-(1+\alpha^{2}/2)\pm\sqrt{E^{2}+(1+\alpha^{2}/2)^{2}-1}.
\end{equation}
This yields two propagating solutions (i.e., real momenta)
\begin{equation}
p_{1,2}=\pm k=\pm\left(-(1+\alpha^{2}/2)+\sqrt{E^{2}+(1+\alpha^{2}/2)^{2}-1}\right)^{1/2},\label{eq:p12}
\end{equation}
and two evanescent ones (imaginary momenta)
\begin{equation}
p_{3,4}=\pm i\kappa=\pm i\left((1+\alpha^{2}/2)+\sqrt{E^{2}+(1+\alpha^{2}/2)^{2}-1}\right)^{1/2}.\label{eq:p34}
\end{equation}
The corresponding spinors are (non-normalized)
\begin{equation}
\Psi_{i}=\begin{bmatrix}\alpha p_{i}\\
E-1-p_{i}^{2}
\end{bmatrix}e^{ip_{i}x},\,i=1,...,4.\label{eq:bspins}
\end{equation}
To find a spatially confined solution, the boundary conditions have
to be satisfied by a linear combination of these four spinors. The
problem can be simplified if one explicitly accounts for the parity
symmetry. Specifically, both Hamiltonian (\ref{eq:H0_Kane_unitless})
and the boundary conditions are symmetric with respect to the parity
transformation, $\hat{\pi}=-\sigma_{z}\hat{P}$, where $\hat{P}\Psi(x)=\Psi(-x)$
is the conventional spatial inversion. Under this condition, it is
convenient to first construct specific-parity linear combinations
of spinors (\ref{eq:bspins}). Then, matching boundary conditions
for such symmetrized spinors involves linear combinations of only
two spinors for each parity. The odd-parity ($\hat{\pi}\Psi=-\Psi$)
spinor can be written as
\begin{align}
\Psi(x)&=A\begin{bmatrix}\alpha k\cos(kx)\\
i\left(E-1-k^{2}\right)\sin(kx)
\end{bmatrix} \nonumber\\
&+B\begin{bmatrix}\alpha\kappa\cosh(\kappa x)\\
i\left(E-1+\kappa^{2}\right)\sinh(\kappa x)
\end{bmatrix},
\end{align}
Coefficients $A$ and $B$ has to found so that $\Psi(+L/2)=0$, and
then $\Psi(-L/2)=0$ is satisfied automatically. Matching this boundary
condition results in the following characteristic equation
\begin{equation}
k\left(E-1+\kappa^{2}\right)\tanh(\kappa L/2)=\kappa\left(E-1-k^{2}\right)\tan(kL/2).\label{eq:gen_bc}
\end{equation}
The even-parity ($\hat{\pi}\Psi=\Psi)$ spinor reads as
\begin{align}
\Psi(x)&=A\begin{bmatrix}\alpha k\sin(kx)\\
-i\left(E-1-k^{2}\right)\cos(kx)
\end{bmatrix}\nonumber\\
&+B\begin{bmatrix}\alpha\kappa\sinh(\kappa x)\\
i\left(E-1+\kappa^{2}\right)\cosh(\kappa x)
\end{bmatrix}.
\end{align}
The characteristic equation, resulting from the boundary conditions, is
\begin{equation}
\kappa\left(E-1-k^{2}\right)\tanh(\kappa L/2)=-k\left(E-1+\kappa^{2}\right)\tan(kL/2).\label{eq:gen_bc_even}
\end{equation}

In general, such transcendental characteristic equations have to be
solved numerically for eigenenergies $E$. However, some general properties
of such solutions can be established right away. First, at $E>1$
both $\left(E-1-k^{2}\right)$ and $\left(E-1+\kappa^{2}\right)$
are always positive. This means that $\tan(kL/2)$ has to be positive
in Eq.~(\ref{eq:gen_bc}) and negative in Eq.~(\ref{eq:gen_bc_even}).
Second, $\tan(kL/2)$ does always vanish at $kL/2=\pi n,\,n=0,1,2,...$
and always diverges to positive or negative infinity when $kL/2$
approaches $\pi(n+1/2)$ from the left or right, respectively. These
two observations imply that Eq.~(\ref{eq:gen_bc}) does always have
one and only one solution within each interval $kL/2\in\pi(n,n+1/2),\,n=0,1,2,...$.
Similarly, Eq.~(\ref{eq:gen_bc}) does always have one and only one
solution within each interval $kL/2\in\pi(n+1/2,2n),\,n=0,1,2,...$.
Therefore, positive-energy states of the spatially confined 1D Kane
model have alternating parities when sorted with respect to their
energies, with the lowest-energy state being the odd one.

Negative-energy solutions ($E<-1$) can be found similarly. However,
It is more convenient to exploit the charge conjugation symmetry of
Hamiltonian (\ref{eq:H0_Kane_unitless}). Specifically, it can be
straightforwardly demonstrated that $CH_{0}C^{\dagger}=-H_{0}$ and
$C\hat{\pi}C^{\dagger}=-\hat{\pi}$, where $C=-i\sigma_{y}$ is the
charge conjugation operator. These identities imply that this transformation
flips the parity and the sign of the energy. More specifically, we
can enumerate all the solutions of the spatially confined 1D Kane
problem according to their energies: $...<E_{-2}<E_{-1}<0<E_{1}<E_{2}<...$.
Then, the charge conjugation symmetry yields: (i) $E_{-n}=-E_{n}$,
(ii) $\Psi_{-n}(x)=C\Psi_{n}(x)$, and (iii) the parity is $\pi_{n}=(-1)^{|n+1/2|-1/2}$.
In particular, the energies of the two band edge states are related
as $E_{-1}=-E_{1}$ with the negative- (positive) energy state being
of the even (odd) parity. In what follows, we consider three limiting
cases, where analytical expressions for energies and wavefunctions
of these band edge states can be obtained explicitly.

\subsection{Weak confinement}

The weak confinement is realized at $L\gg1,\alpha$. At these conditions,
the energy of the positive-energy band edge state is $0<E_{1}-1\ll1$
and Eqs.~(\ref{eq:p12}) and (\ref{eq:p34}) reduce to $k=\sqrt{\frac{E_{1}-1}{1+\alpha^{2}/2}}$
and $\kappa=\sqrt{2(1+\alpha^{2}/2)}$. Then, the solution of Eq.~(\ref{eq:gen_bc})
is $E_{1}=1+(1+\frac{\alpha^{2}}{2})\tilde{p}_{1}^{2}$, where the
effective momentum corresponding to the band edge states is denoted
by $\tilde{p}_{1}=\pi/L$. The corresponding wavefunction is
\begin{equation}
\Psi_{1}(x)=\sqrt{\frac{2}{L}}\begin{bmatrix}\cos(\tilde{p}_{1}x)\\
0
\end{bmatrix}.
\end{equation}
The energy and the wavefunction of the negative-energy band edge state
is obtained using the charge conjugation symmetry as $E_{-1}=-E_{1}$
and 
\begin{equation}
\Psi_{-1}(x)=C\Psi_{1}(x)=\sqrt{\frac{2}{L}}\begin{bmatrix}0\\
\cos(\tilde{p}_{1}x)
\end{bmatrix}.
\end{equation}
The normalized transition intensity (\ref{eq:Ip_Kanec}) can then
be evaluated as
\begin{equation}
\left|\langle\Psi_{1}|f_{1}|\Psi_{-1}\rangle^{2}\right|/\alpha^{2}=1.
\end{equation}

\subsection{Intermediate confinement}

In the intermediate regime, $\alpha^{-1}\ll L\ll\alpha$, the energy
of the lowest positive-energy state is seen in Fig.~\ref{fig:energy_kane_conf}
to be $E_{1}\sim\alpha/L$, so that $1\ll E_{1}\ll\alpha^{2}$. Under
these conditions, Eqs.~(\ref{eq:p12}) and (\ref{eq:p34}) reduce
to $k_{1}=E_{1}/\alpha$ and $\kappa_{1}=\alpha$, respectively. Substituting
these expressions into Eq.~(\ref{eq:gen_bc}) one obtains $\tan\left(E_{1}L/2\alpha\right)=1$,
resulting in 
\begin{equation}
E_{1}=\alpha\tilde{p}_{1}/2,\label{eq:int_conf_Kane}
\end{equation}
where, again, $\tilde{p}_{1}=\pi/L$. The corresponding wavefunction
is 
\begin{equation}
\Psi_{1}(x)=\frac{1}{\sqrt{L}}\begin{bmatrix}\cos(\tilde{p}_{1}x/2)-\sqrt{2}e^{-\alpha L/2}\cosh(\alpha x)\\
i\sin(\tilde{p}_{1}x/2)-i\sqrt{2}e^{-\alpha L/2}\sinh(\alpha x)
\end{bmatrix}.\label{eq:Psi1_int}
\end{equation}
We wish to note here that the naive recipe for performing the 1D spatial
quantization (e.g., how it is done within the parabolic effective
mass approximation) is to evaluate the energy of the spatially confined
state by substituting the effective momentum $\tilde{p}_{1}$ into
the \emph{bulk} dispersion relation, Eq.~(\ref{eq:bulk_disp}). Within
the considered intermediate regime, such a procedure would yield $E_{1}=\alpha\tilde{p}_{1}$,
which differs from the correct expression, Eq.~(\ref{eq:int_conf_Kane}),
by a numerical factor. The inapplicability of the naive quantization
recipe is further emphasized by an observation that even though the
diagonal part of Hamiltonian (\ref{eq:H0_Kane_unitless}) could be
neglected in the continuous case when $\alpha^{-1}\ll p_{1}\ll\alpha$,
such an approximation would be too crude in the spatially quantized
case. Indeed, completely discarding the diagonal part yields a Hamiltonian
matrix which is linear with respect to the momentum operator. For
such a Hamiltonian, there is only a single linearly independent bulk
solution per specific parity, which is not enough to satisfy the boundary
conditions. In Eq.~(\ref{eq:Psi1_int}), the boundary conditions
are satisfied due to the presence of the evanescent tails of the wavefunction,
which are only significant near the edges ($x\approx\pm L/2$). These
tails can only be obtained if Hamiltonian terms proportional to $p^{2}$
are taken into account. That the presence of these evanescent tails
is essential invalidates the naive 1D quantization recipe and results
in the spatial confinement energy, Eq.~(\ref{eq:int_conf_Kane}),
that is a factor of two lower than the naive result, $E_{1}=\alpha\tilde{p}_{1}$. 

The wavefunction for the negative-energy band edge state is obtained
via the charge conjugation transformation
\begin{align}
\Psi_{-1}(x)&=C\Psi_{1}(x) \nonumber\\
&=\frac{1}{L}\begin{bmatrix}-i\sin(\tilde{p}_{1}x/2)+i\sqrt{2}e^{-\alpha L/2}\sinh(\alpha x)\\
\cos(\tilde{p}_{1}x/2)-\sqrt{2}e^{-\alpha L/2}\cosh(\alpha x)
\end{bmatrix}.
\end{align}
The normalized transition intensity can then be evaluated analytically
to yield
\begin{equation}
\left|\langle\Psi_{1}|f_{1}|\Psi_{-1}\rangle\right|^{2}/\alpha^{2}=\frac{4}{\pi^{2}}\approx0.4053.
\end{equation}

\subsection{Strong confinement}

In the limit of the very strong confinement, $L\ll1,\alpha^{-1}$,
one obtains $k^{2}=\kappa^{2}=E_{1}\gg1,\alpha$. The prefactor in
the l.h.s of Eq.~(\ref{eq:gen_bc}) is then much larger than the
prefactor in the r.h.s., and, therefore, the tangent function has
to be very large to compensate for this mismatch. This results $E_{1}=\tilde{p}_{1}^{2}$
for the lowest positive-energy state. The corresponding wavefunction
is
\begin{equation}
\Psi_{1}(x)=A\begin{bmatrix}\alpha\tilde{p}_{1}\cos(\tilde{p}_{1}x)\\
i\frac{\alpha^{2}}{2}\sin(\tilde{p}_{1}x)
\end{bmatrix}+B\begin{bmatrix}\alpha\tilde{p}_{1}\cosh(\tilde{p}_{1}x)\\
2i\tilde{p}_{1}\sinh(\tilde{p}_{1}x)
\end{bmatrix},
\end{equation}
where coefficients $A$ and $B$ are found as follows. The requirement
of vanishing lower component of the spinor at $x=L/2$ yields $B/A=-\alpha\left[4\tilde{p}_{1}^{2}\sinh(\pi/2)\right]^{-1}$.
For this ratio of coefficients, $A\cos(\tilde{p}_{1}x)$ is always
much larger in magnitude than $B\cosh(\tilde{p}_{1}x)$ within the
upper spinor component, so that the latter term can be safely discarded.
The resulting normalized wavefunction reads as
\begin{equation}
\Psi_{1}(x)=\sqrt{\frac{2}{L}}\begin{bmatrix}\cos(\tilde{p}_{1}x)\\
\frac{i\alpha}{2\tilde{p}_{1}}\left[\sin(\tilde{p}_{1}x)-\frac{\sinh(\tilde{p}_{1}x)}{\sinh(\pi/2)}\right]
\end{bmatrix}.
\end{equation}
The wavefunction for the negative-energy band edge state is obtained
using the charge conjugation transformation
\begin{equation}
\Psi_{-1}(x)=C\Psi_{1}(x)=\sqrt{\frac{2}{L}}\begin{bmatrix}-\frac{i\alpha}{2\tilde{p}_{1}}\left[\sin(\tilde{p}_{1}x)-\frac{\sinh(\tilde{p}_{1}x)}{\sinh(\pi/2)}\right]\\
\cos(\tilde{p}_{1}x)
\end{bmatrix}.
\end{equation}
The transition matrix element can then be evaluate analytically to
yield
\begin{equation}
\left|\langle\Psi_{1}|f_{1}|\Psi_{-1}\rangle\right|^{2}/\alpha^{2}=\frac{\left[4\coth(\pi/2)-\pi\right]^{2}}{\pi^{2}}\approx0.1507.
\end{equation}

\section{Kang-Wise problem\label{sec:Kang-Wise}}

Spherically symmetric Dimmock Hamiltonian is given by Eq.~(\ref{eq:Ham_KW}).
Electron and hole size-quantized levels in SN are found as a solution
of a corresponding stationary Schr\"odinger equation subject to boundary
conditions. Specifically, size-quantized levels in a quantum dot (QD)
are found by requiring the wavefunction to be non-singular inside
the QD and to vanish at the QD surface, $\left.\Psi(r)\right|_{|r|=a}=0$,
where $a$ is the QD radius. The standard procedure of finding size-quantized
levels via matching the boundary conditions is as follows \cite{Goupalov-2011-037303}.
First, all the linearly independent eigenfunctions of Hamiltonian
(\ref{eq:Ham_KW}) of specific energy and symmetry are found. Second,
a linear combination of these solutions with unknown coefficients
is required to vanish at the QD surface - this condition yields a
secular equation. Solving this secular equation results in the energies
of the size-quantized states. Below we demonstrate how to employ symmetry
considerations to construct the exact ansatz for a wavefunction and
thus simplify the solution of the problem.

The symmetries of the Hamiltonian (\ref{eq:Ham_KW}) and of the boundary
conditions are given by the following operators:
\begin{enumerate}
\item Parity 
\begin{equation}
\Pi_{4}=\begin{bmatrix}-\pi I_{2} & 0\\
0 & \pi I_{2}
\end{bmatrix},\label{eq:Par_op}
\end{equation}
where $\pi$ is a coordinate inversion operator (${\bf r}\rightarrow-{\bf r}$).
The global sign of the operator is chosen to represent the parity
of the total microscopic wavefunction, since the conduction band Bloch
function is odd, and the valence band Bloch function is even \cite{Kang-1997-1632}.
It is clear, however, that as long as $k\cdot p$ Hamiltonian is specified,
parity is its internal property and the original parity of the Bloch
functions are irrelevant. For example, even if a certain material
does not have any inversion symmetry, but a corresponding effective
Hamiltonian could be written as Eq.~(\ref{eq:Ham_KW}), then the
parity operator (\ref{eq:Par_op}) could be associated with it.
\item Total angular momentum and its projection
\begin{equation}
J_{4,z}=\begin{bmatrix}J_{z} & 0\\
0 & J_{z}
\end{bmatrix},\,\,J_{4}^{2}=\begin{bmatrix}J^{2} & 0\\
0 & J^{2}
\end{bmatrix},
\end{equation}
Here, the operator of the total angular momentum is given by ${\bf J}={\bf L}I_{2}+\boldsymbol{\sigma}/2$,
where ${\bf L}$ is the operator of the orbital momentum.
\item Square of the momentum
\begin{equation}
p_{4}^{2}=p^{2}I_{4},
\end{equation}
where $I_{4}$ is the $4\times4$ unit matrix.
\end{enumerate}
All these operators commute with each other and the Hamiltonian,
so it is possible to find wavefunctions of Hamiltonian (\ref{eq:Ham_KW}),
which would satisfy all the symmetries above. Square of the momentum
is not a conventional symmetry operator, but it can be straightforwardly
shown to commute with the Hamiltonian and will be of use in what follows.
Importantly, all the symmetry operators are block-diagonal, i.e.,
all these symmetries are applicable not only to the entire four-spinors,
but also to the upper or lower bi-spinor separately. Specifically,
the fact that $J_{z}$, $J^{2}$ and $\pi I_{2}$ have to be conserved
for an upper and lower bi-spinor independently, results in that the
angular part of a bi-spinor of a specific parity can be represented
as a spin spherical harmonic \cite{Edmonds-1957-angular}
\begin{equation}
\chi_{jlm}(\Omega)=\begin{bmatrix}C(l,m-1/2;1/2,1/2;j,m)Y_{l,m-1/2}(\Omega)\\
C(l,m+1/2;1/2,-1/2;j,m)Y_{l,m+1/2}(\Omega)
\end{bmatrix},
\end{equation}
where $j$, $l$ and $m$ are the total angular momentum, orbital
momentum of the coordinate part of the wavefunction, and the projection
of the total angular momentum, respectively. Clebsh-Gordan coefficients
are denoted by $C(j_{1},m_{1};j_{2},m_{2};j,m)$. The parity of this
bi-spinor with respect to ${\bf r}\rightarrow-{\bf r}$ is given by
$(-1)^{l}$. For a given $j$ and $m$, there are only two possible
bi-spinors with orbital momenta $l=j\pm1/2$.

To obtain a full bi-spinor, one has to multiply its angular part by
a radial part, $f_{l}(r)$. Since bi-spinors have to be eigenfunctions
of operator $p^{2}I_{2}$, we have
\begin{align}
p^{2}\chi_{jlm}(\Omega)f_{l}(r)&=-\chi_{jlm}(\Omega)\left[\frac{\partial^{2}}{\partial r^{2}}+\frac{2}{r}\frac{\partial}{\partial r}-\frac{l(l+1)}{r^{2}}\right]f_{l}(r) \nonumber \\
&=\kappa^{2}\chi_{jlm}(\Omega)f_{l}(r),
\end{align}
where $\kappa^{2}$ denotes the eigenvalue corresponding to operator
$p^{2}$. The solution of this equation is the spherical Bessel function
of the first kind, $f_{l}(r)=j_{l}(\kappa r)$ \cite{Abramowitz-Stegun-1965}.
Spherical Bessel function of the second kind is also the solution,
but it is not suitable for the expansion of the final wavefunction
since it is singular at $r=0$. 

Operator $\boldsymbol{\sigma}\cdot{\bf p}$ flips the parity of the
bi-spinor, does not change $j$ or $m$, and commutes with $p^{2}$.
Therefore, operator $\boldsymbol{\sigma}\cdot{\bf p}$ can only transform
a bi-spinor with quantum numbers $(j,l,m)$ to the one with quantum
numbers $(j,l',m)$, where $l'=2j-l$. It is thus expected that the
radial part of operator $\boldsymbol{\sigma}\cdot{\bf p}$ has to
be a raising or lowering operator for the spherical Bessel functions.
Indeed, if one combines $j$ and $l$ into a single quantum number
$k=\mp(j+1/2)$ for $j=l\pm1/2$, then spherical spinors can be written
as $\chi_{km}(\Omega)$ and one has \cite{Johnson-2004-120}
\begin{equation}
\boldsymbol{\sigma}\cdot{\bf p}\chi_{km}(\Omega)=i\left(\frac{\partial}{\partial r}+\frac{k+1}{r}\right)\chi_{-km}(\Omega).\label{eq:sigma_p}
\end{equation}
The radial operator in parentheses becomes $A_{l}^{-}$ for positive
$k$ ($k=l$) and $-A_{l}^{+}$ for negative $k$ ($k=-l-1$), where
the raising and lowering operators for the spherical Bessel functions
are defined as
\begin{gather}
A_{l}^{-}j_{l}(r)=\left(\frac{\partial}{\partial r}+\frac{l+1}{r}\right)j_{l}(r)=j_{l-1}(r),\nonumber \\
A_{l}^{+}j_{l}(r)=-\left(\frac{\partial}{\partial r}-\frac{l}{r}\right)j_{l}(r)=-j_{l+1}(r).\label{eq:bessel_lower_raise}
\end{gather}
Eqs.~(\ref{eq:sigma_p}) and (\ref{eq:bessel_lower_raise}) can be
combined into
\begin{equation}
\boldsymbol{\sigma}\cdot{\bf p}F_{km}({\bf r})=i\kappa{\rm sgn}(k)F_{-km}({\bf r}),
\end{equation}
where $F_{km}({\bf r})=\chi_{km}(\Omega)j_{l}(\kappa r)$ is the full
bi-spinor. Then, if we construct the four-spinor as
\begin{equation}
\Psi({\bf r)=}\begin{bmatrix}F_{km}({\bf r)}\\
i{\rm sgn}(k)F_{-km}({\bf r)}
\end{bmatrix},
\end{equation}
it is clear that acting onto such a spinor by the Dimmock Hamiltonian
(\ref{eq:Ham_KW}) effectively transforms this Hamiltonian into a
$2\times2$ $c$-number matrix
\begin{equation}
H_{0}\rightarrow\begin{bmatrix}\left(\frac{E_{g}}{2}+\frac{\kappa^{2}}{2m_{c}}\right) & P\kappa\\
P\kappa & -\left(\frac{E_{g}}{2}+\frac{\kappa^{2}}{2m_{v}}\right)
\end{bmatrix},
\end{equation}
so that the energy and $\kappa$ are related via
\begin{equation}
\left[\frac{E_{g}}{2}+\frac{\kappa^{2}}{2m_{c}}-E\right]\left[\frac{E_{g}}{2}+\frac{\kappa^{2}}{2m_{v}}+E\right]+P^{2}\kappa^{2}=0.
\end{equation}
For each specific energy this equation has four different solutions
$\kappa$. Each pair $(\kappa,-\kappa)$ produces only a single linearly-independent
solutions, so only two $\kappa$'s out of the four are used to construct
two linearly independent four-spinors. Each component of the linear
combination of these two spinors has to vanish at the QD surface,
which produces a homogeneous system of two equations with two unknowns.
A secular equation has to be satisfied for this system to have non-trivial
solution, thus resulting in the size-quantization condition. The explicit
expression for the secular equation can be found in Ref.~\cite{Kang-1997-1632}.
The solution of the Kang-Wise problem for the cylindrical symmetry
is provided in Ref.~\cite{Goupalov-2011-037303}.

\section*{References}


\end{document}